# Interplanetary shocks lacking type II radio bursts

**N. Gopalswamy[1], H. Xie[2], P. Mäkelä[2], S. Akiyama[2], S. Yashiro[3], M. L. Kaiser[1],**

**R. A. Howard[4] and J.-L. Bougeret[5]**

[1]NASA Goddard Space Flight Center, Greenbelt, Maryland

[2] The Catholic University of America, Washington, DC

[3]Interferometrics, Herndon, VA

[4]Naval Research Laboratory, Washington, DC

[5]Paris Observatory, Meudon, France

E- mail: nat.gopalswamy@nasa.gov

**Abstract.** We report on the radio-emission characteristics of 222 interplanetary (IP) shocks detected by spacecraft at Sun-Earth L1 during solar cycle 23 (1996 to 2006, inclusive). A surprisingly large fraction of the IP shocks (~34%) was radio quiet (i.e., the shocks lacked type II radio bursts). We examined the properties of coronal mass ejections (CMEs) and soft X-ray flares associated with such radio-quiet (RQ) shocks and compared them with those of the radio-loud (RL) shocks. The CMEs associated with the RQ shocks were generally slow (average speed ~535 km/s) and only ~40% of the CMEs were halos. The corresponding numbers for CMEs associated with RL shocks were 1237 km/s and 72%, respectively. Thus, the CME kinetic energy seems to be the deciding factor in the radio-emission properties of shocks. The lower kinetic energy of CMEs associated with RQ shocks is also suggested by the lower peak soft X-ray flux of the associated flares (C3.4 vs. M4.7 for RL shocks). CMEs associated with RQ CMEs were generally accelerating within the coronagraph field of view (average acceleration ~+6.8 $ms^{-2}$), while those associated with RL shocks were decelerating (average acceleration ~ - 3.5 $ms^{-2}$). This suggests that many of the RQ shocks formed at large distances from the Sun, typically beyond 10 Rs, consistent with the absence of metric and decameter – hectometric (DH) type II radio bursts. A small fraction of RL shocks had type II radio emission solely in the kilometric (km) wavelength domain. Interestingly, the kinematics of the CMEs associated with the km type II bursts is similar to those of RQ shocks, except that the former are slightly more energetic. Comparison of the shock Mach numbers at 1 AU shows that the RQ shocks are mostly subcritical, suggesting that they were not efficient in accelerating electrons. The Mach number values also indicate that most of these are quasi-perpendicular shocks. The radio-quietness is predominant in the rise phase and decreases through the maximum and declining phases of solar cycle 23. About 18% of the IP shocks do not have discernible ejecta behind them. These shocks are due to CMEs moving at large angles from the Sun-Earth line and hence are not blast waves. The solar sources of the shock-driving CMEs follow the sunspot butterfly diagram, consistent with the higher-energy requirement for driving shocks.

## 1. Introduction

Interplanetary (IP) shocks driven by coronal mass ejections (CMEs) are indicative of powerful eruptions on the Sun that accelerate particles to very high energies. The shocks can be inferred from type II radio bursts they produce in the corona and IP medium. Every large solar energetic particle (SEP) event is associated with a type II radio burst (Gopalswamy 2003; Cliver *et al* 2004), which is used as a strong evidence for particle acceleration by shocks (Gosling 1993; Reames 1999). When the shocks arrive at Earth, they can be identified from the energetic storm particle (ESP) events (Bryant *et al* 1962; Rao *et al* 1967). Although ESP events are thought to indicate acceleration of protons to tens of MeV, it is expected that the shocks near the Sun accelerate particles to GeV energies observed in ground level enhancements (GLEs) in SEP events (McCracken *et al* 2008). This is because the shocks are likely to be stronger near the Sun where the driving CMEs attain their maximum speed before decelerating due to the drag force exerted by the ambient medium (see e.g., Vrsnak 2001; Gopalswamy *et al* 2001b). CME-driven shocks have been found to be very efficient particle accelerators in that ~10% CME kinetic energy is converted into SEP kinetic energy (see e.g., Mewaldt 2006). Shocks arriving at Earth also compress the magnetosphere causing the storm sudden commencement (SSC), which may be followed by a geomagnetic storm if the shock sheath and/or the driving IP CME (ICME) contains south-pointing magnetic field (see e.g., Tsurutani *et al* 1988; Gopalswamy *et al* 2008a). Shocks can also be directly detected in situ in the solar wind data as a discontinuous jump in density, temperature, flow speed, and magnetic field.

As noted above, shocks detected at 1 AU have already evolved for a day or more in the IP medium, so they do not reflect their initial properties near the Sun. Since type II radio bursts are the earliest indicators of shocks, one should look at the radio emission characteristics near the Sun and in the IP medium (see e.g., Gopalswamy *et al* 2008b, c). The driving CMEs are readily imaged near the Sun, so the CME properties are useful in understanding the expected shock characteristics near the Sun. Direct detection of shocks from white-light coronagraphic observations is possible only occasionally (Sheeley *et al* 2000; Vourlidas *et al* 2003) so we need to rely on a combination of radio burst observations and white light observations of the driving CMEs. The strength of the shock is determined not only by the driver speed, but also by the characteristic speed (such as the Alfven speed) of the ambient medium. Although the Alfven speed at 1 AU is rather small (~75 km/s), it can be higher by more than an order of magnitude near the Sun (Krogulec *et al* 1994; Mann *et al* 1999; Gopalswamy *et al* 2001a; Mann *et al* 2003; Gopalswamy *et al* 2008b, c). One of the recent findings is that while all type II bursts are indicative of shocks, some shocks are not associated with type II radio bursts near the Sun or in the IP medium (Gopalswamy 2008). Lack of type II bursts implies lack of electron acceleration because type II bursts are caused by electrons in the energy range 0.2 - 10 keV accelerated at the shock front (see e.g., Bale *et al* 1999; Knock *et al* 2001; Mann and Klassen 2005). The energetic electrons are unstable to Langmuir waves, which get converted into radio emission at the local plasma frequency and its harmonic (see e.g., Nelson and Melrose 1985 for more details). Thus, type II radio burst contains information on both the shock and the ambient medium in which the shock propagates. This paper is aimed at understanding the IP shocks lacking type II radio bursts using properties of the associated CMEs and those of the ambient medium. To this end, we make use of the large number of IP shocks detected during cycle 23 and the data on CMEs and radio bursts available throughout the solar cycle. There have been a few recent investigations on interplanetary shocks (Oh *et al* 2007, Howard and Tappin 2005), but these did not focus on type II radio emission. Our investigation is focused on the solar source (CME and flare) and radio emission properties of interplanetary shocks to find out what distinguishes the shocks that do and do not produce type II radio bursts.

The rest of the paper is organized as follows: Section 2 describes the data on shocks and the associated phenomena such as ICMEs, CMEs, flares, and type II radio bursts. Section 3 compares the properties of shocks (nature of the ICMEs, shock speed, and Alfvenic Mach number) distinguishing those that do and do not produce type II radio bursts. Section 4 provides detailed information on the CMEs that drive the shocks emphasizing on the kinematics, associated soft X-ray flares, and the eruption locations on the solar disk. Section 5 discusses the results and the last section (6) summarizes the findings.

## 2. Data and analysis

The starting point of this paper is the set of IP shocks detected near Earth by one or more of the following spacecraft: the Solar and Heliospheric Observatory (SOHO), the Advanced Composition Explorer (ACE), and Wind. From the compiled list, we eliminated the shocks known to be driven by Corotating Interaction Regions (CIRs). For each of the remaining 222 shocks, we identified the following: (i) the solar wind driver (magnetic cloud or ejecta) using ACE and Wind data, (ii) the CME driver at the Sun from SOHO's Large Angle and Spectrometric Coronagraph (LASCO, Brueckner *et al* 1995), (iii) the flare size and location from the Solar Geophysical Data (SGD, ftp://ftp.ngdc.noaa.gov/STP/SOLAR_DATA/SOLAR_FLARES/XRAY_FLARES), and (iv) type II radio burst association from the Solar Geophysical data SGD and the Radio and Plasma Wave Experiment (WAVES, Bougeret *et al* 1995) on board Wind. Based on the radio observations, we define shocks lacking detectable type II radio emission as radio quiet (RQ), similar to the criterion used by Gopalswamy *et al* (2008b, c) in defining RQ fast and wide CMEs. Note that radio quietness refers just to the lack of type II bursts, and not to other types such as type III bursts.

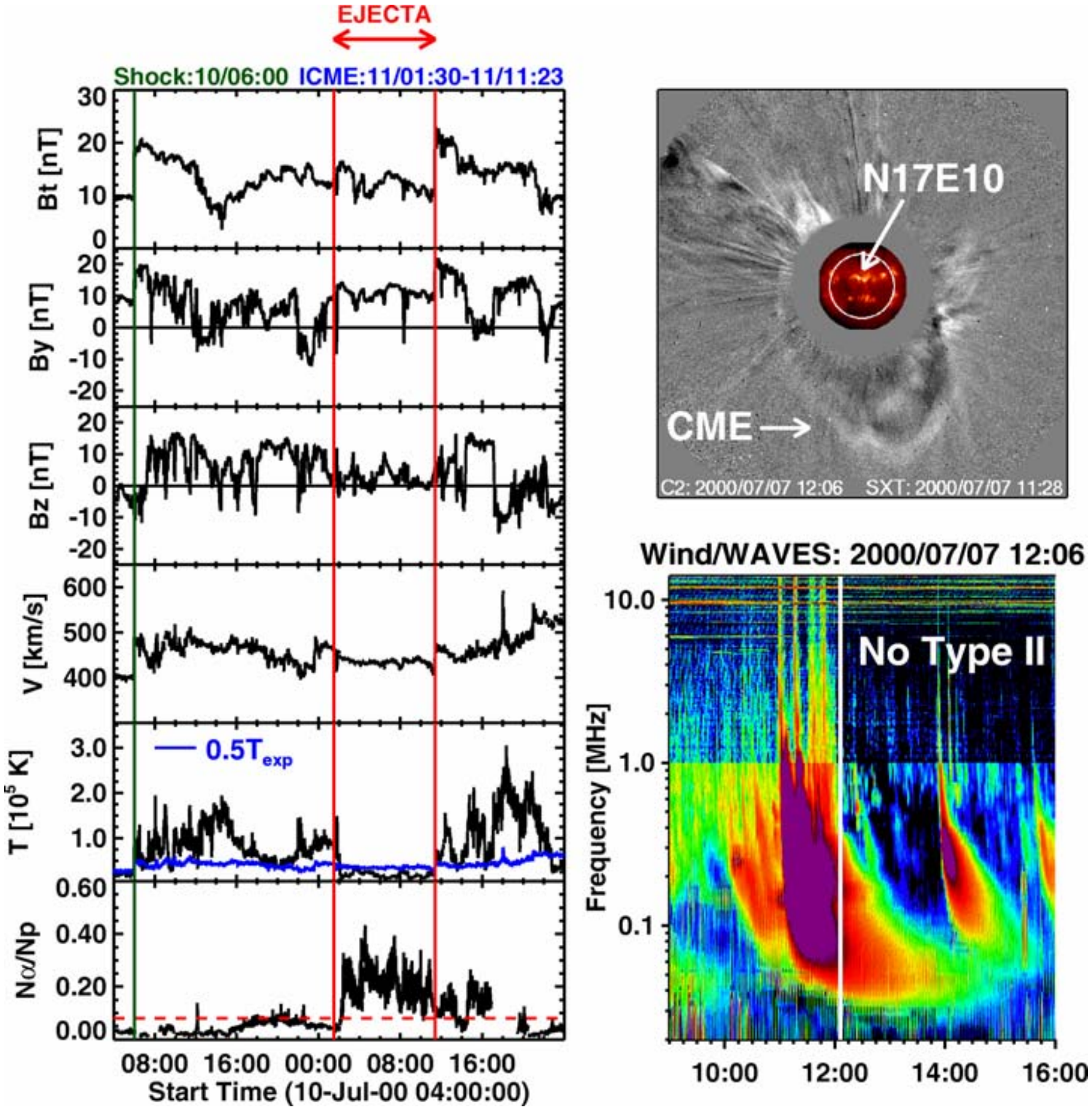

Figure 1. A radio-quiet shock (left), the associated CME (top right) and radio emission (bottom right). The solar wind magnetic field magnitude (Bt) and the components By and Bz, the flow speed (V), the proton temperature (T) and the alpha to proton density ratio (Nα/Np) are plotted. The ejecta is indicated by the interval that shows proton temperature depression below half the expected solar wind temperature (Texp). The enhanced Nα/Np also roughly coincides with the interval of temperature depression. The enhanced temperature to the left of the ejecta corresponds to the shock sheath and the one to the right of the ejecta corresponds to the CIR due to the high-speed stream. The white-light CME at 12:06 UT from SOHO/LASCO has been identified to be the driver of the IP shock. The coronagraph image has a Soft X-ray Telescope (SXT) image overlaid to show the solar source. The SXT image shows the post-eruption arcade at N17E10 and a coronal hole immediately to the east of eruption region. The Wind/WAVES radio dynamic spectrum shows intense type III bursts from the eruption, but no type II burst.

Once the IP shock is identified, we can search for the ICME signature behind the shock. Among the many solar wind signatures of ICMEs (see, e.g., Neugebauer and Goldstein 1997 and references therein), we take the interval of proton temperature depression as the ICME interval. Figure 1 shows a RQ IP shock, which arrived at the Sun-Earth L1 point on 2000 July 10 at 06:00 UT. Immediately after the shock, the solar wind proton temperature increases well above the expected solar wind proton temperature (Texp, see Lopez and Freeman 1986), which is typical of shock sheaths. During a 10-h interval starting at 01:30 UT on 2000 July 11, we see that the proton temperature dips well below 0.5Texp, which is typical of ICMEs. Although the total field (Bt) in the ICME is enhanced above the quiet solar wind value, the field does not rotate smoothly in the Y or Z directions, so this is not a magnetic cloud (MC). We refer to ICMEs without MC signature as non-cloud ejecta, or simply ejecta (EJ). The ratio of alpha-particle density to the proton density in the solar wind (Nα/Np) – another ICME signature - is clearly enhanced during the ejecta interval (see figure 1). The solar wind speed increases right after the ejecta because of a high speed stream, reaching a peak speed of ~550 km/s.

The CME associated with the shock in figure 1 has been identified as the halo CME on 2000 July 7 at 10:26 UT with a sky-plane speed of 453 km/s within the LASCO field of view. This CME originated from a complex active region near the disk center (N17E10) associated with a C5.6 soft X-ray flare. We refer to the heliographic coordinates of the flare location (N117E10) as the solar source of the CME that produced the shock in question. There were a few other wide CMEs that occurred after the July 7 CME, but their source locations were not favorable for Earth arrival. The July 7 CME can be identified as the solar source because of another reason: there was a low-latitude coronal hole immediately to the east of the eruption, which was responsible for the high-speed stream that was immediately behind the ejecta observed at 1 AU. A snapshot of the LASCO CME at 12:06 UT is shown in figure 1 with a Yohkoh soft X-ray telescope (SXT) image superposed showing the post-eruption arcade. The height-time measurements of the CME fit to a second order polynomial, indicating an acceleration of ~10 $ms^{-2}$. The radio dynamic spectrum in figure 1 from Wind/WAVES shows that there was a group of type III bursts associated with the eruption, but no type II burst. From the list of metric type II bursts available from the SGD publications, we confirmed that there was no metric type II burst. Thus the IP shock was not associated with type II radio emission anywhere in the spatial domain between the Sun and Earth. This means that the shock was not able to accelerate enough nonthermal electrons to produce a type II radio burst. If the shock were radio loud, one would see a type II burst as a slanted feature to the right of the type III burst group. Examples of RL shock/CME have been published in several places before, so we have not given one here (see, e.g., Gopalswamy *et al* 2001b; Gopalswamy 2004a; Gopalswamy *et al* 2005; 2009a). Using the method described above, we identified the solar sources, the type II burst association, and the IP drivers of all the 222 shocks.

The IP shocks are listed in the electronic supplement, with a sample of 10 shocks given in table 1. Table 1 contains information on the shocks, the associated ejecta, the solar source (CME, flare), and type II radio emission (metric and IP). For shocks and ejecta, the starting time and speed at 1 AU are given along with the shock observing spacecraft (A = ACE, W = Wind, S = SOHO). For each shock, the computed Alfvenic Mach number ($M_A$) is also listed. For CMEs, the central position angle (CPA), apparent angular width (W in degrees), and the sky-plane speed (V in km/s) are given. Information on CMEs was extracted from the CME catalog (http://cdaw.gsfc.nasa.gov/CME-list; Yashiro *et al* 2004; Gopalswamy *et al* 2009b). The solar source identification of IP shocks is somewhat difficult, but we use the general procedure employed in previous studies (Gopalswamy *et al* 2000a; 2007; 2009a): For each shock, we compile all the CMEs detected by SOHO/LASCO over an interval of 1-5 days preceding the shock arrival and choose the best candidate that occurs on the front side of the Sun and is compatible with the observed in-situ speed. Once the CME is identified, we then determine the solar source as the heliographic location of the associated flare as derived from SGD. When H-alpha flare information is not available, we identified the solar source using one or more of the following signatures: the post-eruption arcade, coronal dimming, or large-scale wave disturbances observed in SOHO's Extreme-ultraviolet Imaging Telescope (EIT, Delaboudinère, *et al* 1995). Determination of the radio quietness is based on checking the type II burst data available on line

ftp://ftp.ngdc.noaa.gov/STP/SOLAR_DATA/SOLAR_RADIO/SPECTRAL/Type_II_1994-2007) at the National Geophysical Data Center (NGDC) for metric type II bursts and the Wind/WAVES data also made available online (http://cdaw.gsfc.nasa.gov/CME_list/radio/waves_type2.html). In the metric domain the type II burst information is given as follows: 0 – no type II burst and 1 – metric type II burst is observed and the starting time is indicated. Similarly, in the longer wavelength domain, 0 – no type II burst, 1 – type II burst in the 1-14 MHz range; 2 – type II burst at frequencies below 1 MHz; 3 – type II bursts at both high and low frequencies (14 MHz to below 1 MHz). Thus 0 – 0 combinations in columns 19 and 21 indicate a RQ shock. All other shocks are RL. The 1 – 0, combination indicates shocks associated with purely metric type II bursts. In these cases, the radio emission was detected only when the shock was within 2 Rs from the Sun. Beyond this distance, the shock was radio quiet. The combination 0 – 2 indicates shocks associated with purely kilometric type II bursts. In these cases, the radio emission is produced far from the Sun (typically beyond 10 Rs). It was not possible to whether the shock of 2001 April 7 at 17:58 UT RQ or RL because of a data gap in the radio observations. In all, there were 76 RQ shocks and 145 RL shocks, which are the focus of our study in this paper.

**3. Shock properties**

Over a period of 11 years, only 230 CME-driven shocks were detected. This corresponds to an average annual rate of ~21 shocks per year, making them a rare phenomenon compared to the number of CMEs and major flares detected over the same period. The actual number of shocks is likely to be higher because we did not include the shocks detected during major SOHO data gaps (for lack of CME association) and a few shocks which are likely to be slow or intermediate type shocks. In four cases, the shocks were identified, but CME data gaps prevented us from obtaining the solar source properties. In one case, there was radio data gap, so we were not able to determine whether the shock is radio quiet or not. In three cases, the solar source was identified, but the CMEs could not be measured, so we exclude these shocks from statistics. We use the remaining set of 222 shocks in the present study (76 RQ, 145 RL, and one with radio data gap). In this section we provide an overview of the shock properties. The number of shocks per day (see table 2) is similar to that reported by Sheeley *et al* (1985) during the maximum and declining phases of cycle 21. The rate during the maximum phase (~0.1 per day) is twice that of the rise and declining phases (~0.04). For individual years, the daily occurrence rate varied from ~0.01 (year 1996) to 0.12 (year 2001). The numbers of RQ and RL shocks in the present work are sufficiently large so that meaningful statistical results can be obtained.

Most of the shocks were followed by ICMEs, which are either MCs or non-cloud ejecta. Only those ICMEs that can be fit to a flux ropes are included under MCs (see e.g., Lepping *et al* 2006). We cross-checked the MC events with the list provided online at the Wind Magnetic Filed Investigation (MFI) web site (http://lepmfi.gsfc.nasa.gov/mfi/mag_cloud_S1.html). An updated list with the solar source identification can be found in Gopalswamy *et al* (2009e). A significant number (42 out of 222 or 19%) of shocks were not followed by a discernible driver. We refer to these as "driverless" shocks; they do have drivers, but the drivers are not intercepted by the observing spacecraft (Gopalswamy *et al* 2001a). Of the remaining 180 shocks, 57 (or 32%) were driven by MCs and the remaining 123 (or 68%) were driven by non-cloud ejecta. In this respect, our result differs substantially from that of Oh *et al* (2007), who associated the majority of IP shocks with MCs rather than with non-cloud ejecta. According to the classical definition, ICMEs with enhanced magnetic field strength, smooth rotation of the field component perpendicular to the Sun-Earth line, and low proton temperature are the MCs (Burlaga *et al* 1981). The definition of MCs used by Oh *et al* (2007) seems to differ substantially from the classical definition of MCs and hence might have contributed to the discrepancy.

### *3. 1 Shock and ejecta speeds*

Figure 2 shows the distribution of shock and ejecta speeds measured at 1 AU. Both ejecta and shocks are generally faster than the slow solar wind. The average speed of the ejecta is smaller than that of the shocks in all the three sets (RQ, RL, and all shocks). The RQ shocks and their ejecta have below-average speeds, while the RL shocks and their ejecta have above-average speeds. The average speed of RQ shocks is ~34% smaller than that of the RL shocks, with a similar difference for ejecta (~28%). Thus there is a clear indication that the RQ shocks are generally weaker than the RL shocks. The median speeds confirm the same tendency that RQ shocks and their ejecta are generally weak compared to the RL populations.

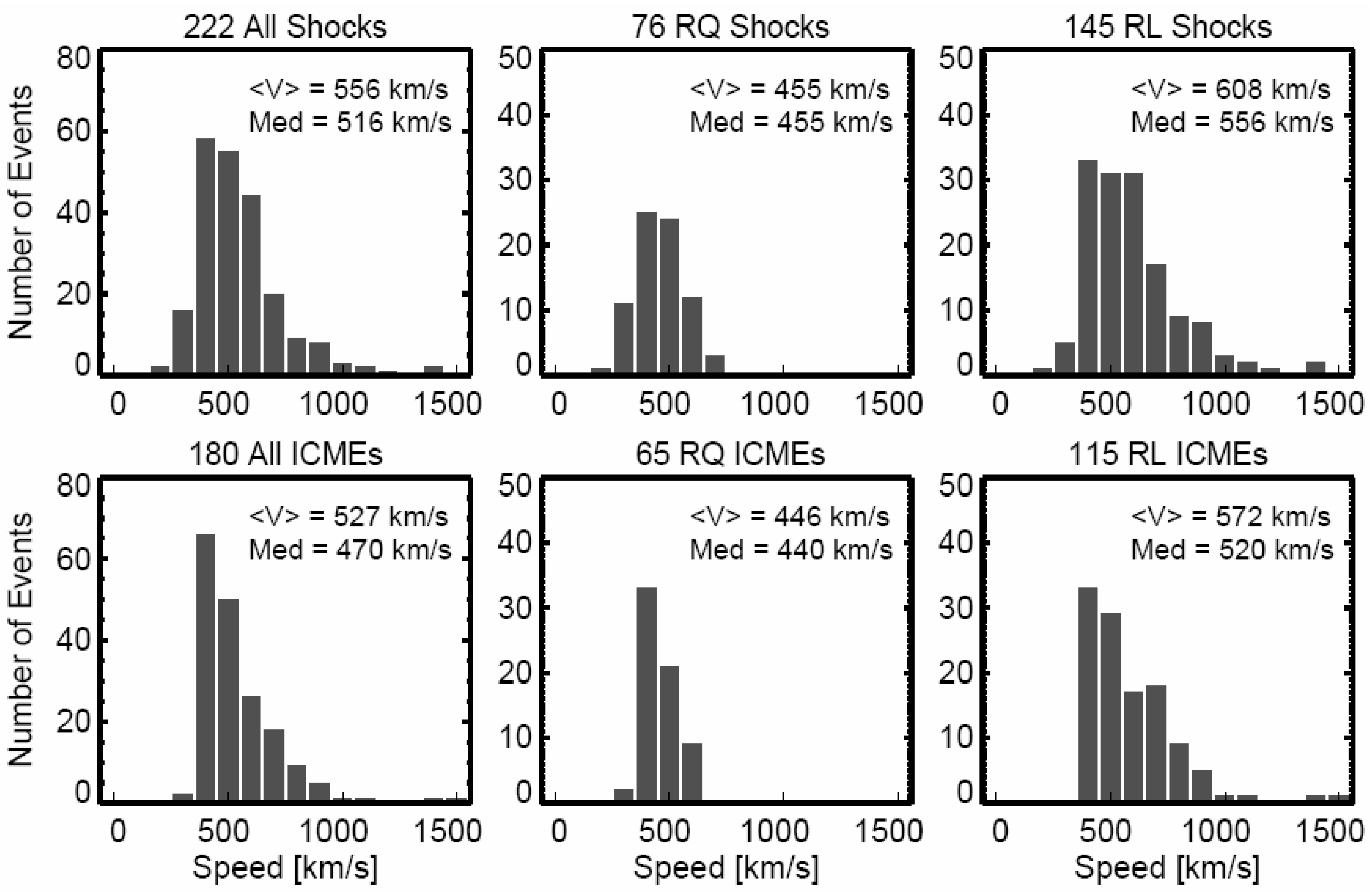

Figure 2. Distributions of shock (top panels) and ejecta (bottom panels) speeds for all shocks (left), RQ shocks (middle) and RL shocks (right). The average (<V>) and median (Med) speeds in each case are noted on the plots. The bin size is 100 km/s. The 500 km/s bin contains events with speeds from 450 km/s to 550 km/s.

In order to further quantify the strength of the shocks, we determined the Alfvenic Mach number ($M_A$) for each shock from in situ data. The Mach numbers were obtained from one of the following three methods: (1) from the compilation of shock properties made available online by J. C. Kasper (http://www.cfa.harvard.edu/shocks/) we extracted the Mach numbers for 151 shocks (Mach numbers are average values computed using eight different methods); (2) for 64 events with good solar wind data but not listed in the above web site, we used the Shock and Discontinuities Analysis Tool (SDAT). SDAT uses an extension of the Viñas-Scudder (1986) analysis method based on the Rankine-Hugoniot conservation equations; (3) for 15 events with low data quality we used the following approximate method to estimate $M_A$**.** Assuming that the shock normal and speed are along the radial direction, the shock speed ($V_{sh}$) and $M_A$ can be obtained from the mass conservation relation:

$(V_d - V_{sh})N_d = (V_u - V_{sh})N_u$, (1)

where the subscripts d and u to the solar wind speed (V) and density (N) indicate the downstream and upstream values, respectively. Inputting $V_d$ and $V_u$ into Equation 1, we get $V_{sh}$. From the upstream magnetic field ($B_u$) and $N_u$, we obtain the upstream Alfven speed $V_A$ and hence $M_A = (V_{sh} – V_u)/ V_A$. We randomly selected 8 events in J. C. Kasper list and computed $M_A$ using the SDAT and approximate methods. $M_A$ differs by 0.55 between methods (1) and (2) and by 0.73 between (1) and (3). Similarly, the estimated shock speed differs by 38 km/s between (1) and (2), and 82 km/s between (1) and (3).

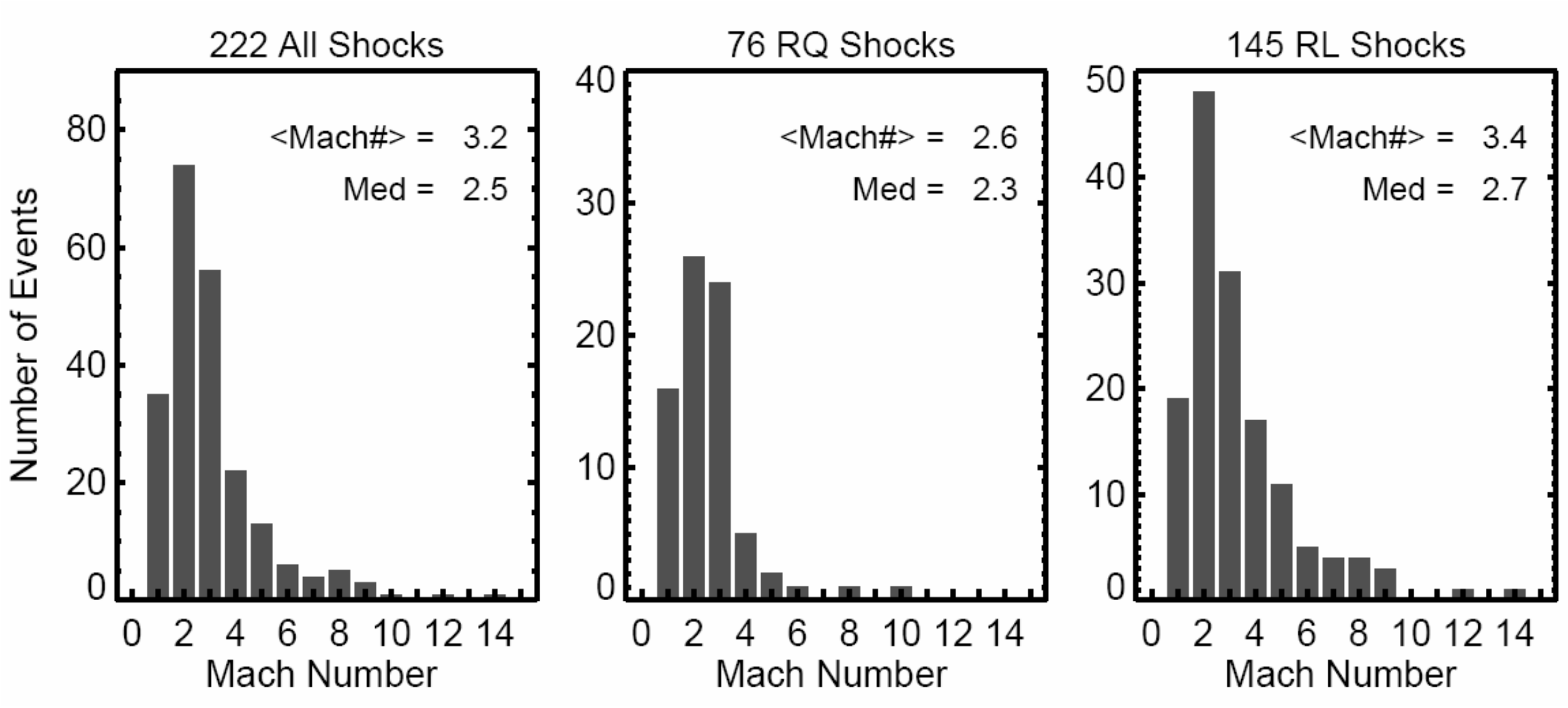

Figure 3. Distribution of Alfvenic Mach numbers for all the shocks (left), RQ shocks (middle), and RL shocks (right). The average (<Mach#>) and median (Med) values of the distributions are given on the plots. The bin size is 1. The bin labeled 2 contains events with Mach numbers from 1.5 to 2.5.

Figure 3 shows the distribution of $M_A$ for all the shocks and the RQ and RL subsets. The average $M_A$ for all the shocks is ~3.2. For RL shocks, $M_A$ = 3.4, which is ~31% higher than that of the RQ shocks ($M_A$ = 2.6). The Mach number difference reflects the CME speed difference for RQ and RL shocks and confirms that the RQ shocks are weaker. Since the Mach number distributions are not symmetric, we have also given the median values. Although the median $M_A$ values of the RL and RQ shocks are above and below the median $M_A$ for all shocks, we see that the difference is not very high. We shall discuss the reason for this in the next section.

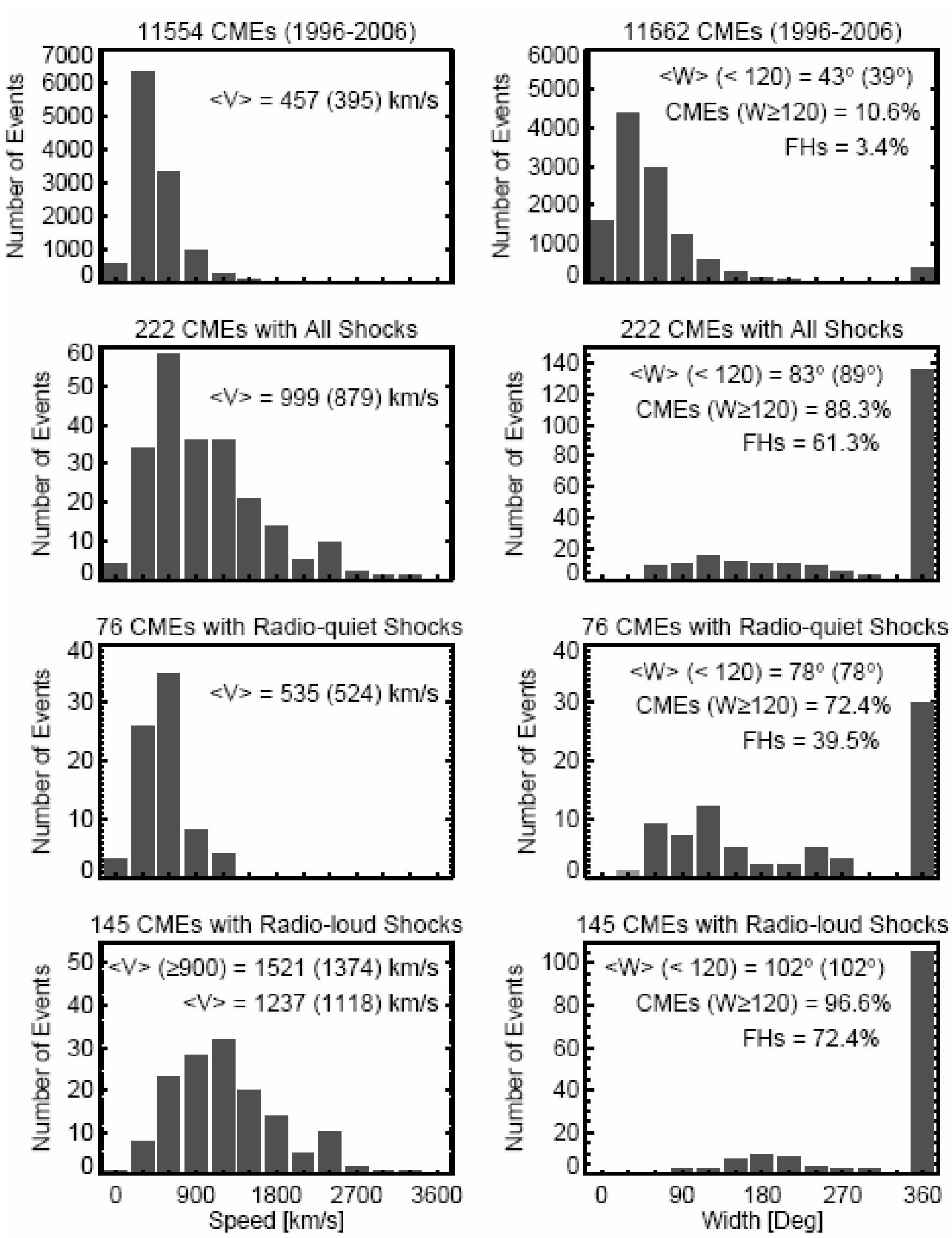

Figure 4. Speed (V, left) and width (W, right) distributions of CMEs associated with various shock populations compared with those of the general population of CMEs (1996 – 2006). The average values of the distributions are given on the plots with the median values in parentheses. In the bottom left-hand panel, the average speed (1521 km/s) of fast CMEs (speed ≥900 km/s) is also shown. The bin size of the speed distribution is 300 km/s. The 900 km/s bin contains events with speeds from 750 km/s to 1050 km/s. In the width column, the average

values correspond to non-halo CMEs (W < 120$^{o}$). Fraction of partial halos (W ≥ 120$^{o}$) and full halos (FHs, W = 360$^{o}$) are also shown. The bin size of the width distribution is 30$^{o}$ (e.g., the 90$^{o}$ bin contains events with widths from 75$^{o}$ to 105$^{o}$).

## 4. CME properties

Each IP shock in table 1 is associated with a white-light CME observed near the Sun. Shocks take anywhere from ~19 h to more than 3 days to arrive at Earth. The best possible candidate CME is selected from the list of all CMEs that occurred 1-4 days prior to the shock arrival at Earth. The number of shock-driving CMEs (222) is rather small compared to the >11,000 CMEs detected during the study period (1996 - 2006). Even if only half of these CMEs are frontsided, it is clear that only a few percent of the CMEs produce detectable shocks near Earth. It is worth comparing the properties of the source CMEs and flares to gain some insight into the reason for the lack of type II burst association for some shock-driving CMEs. In particular, we compare the CME speed, angular width, X-ray flare size, and the source distributions of RQ and RL shocks.

### *4.1CME Speed distributions*

Figure 4 shows the speed and width distributions of CMEs associated with all, RQ, and RL shocks. The speeds and widths are in the sky plane and no attempt was made to correct for projection effects. The average speed (535 km/s) of CMEs associated with RQ shocks is only slightly greater than that of the general population of CMEs (457 km/s). The average CME speed of RL shocks is a factor of ~ 2 higher than that of the RQ shocks (1237 km/s vs. 535 km/s). Note that the difference between the two populations is more pronounced at the Sun than at 1 AU (see figures 2 and 3). This is expected because of the momentum exchange between CMEs and the solar wind during the interplanetary passage (see section 4.3). Among the RL shocks, 11 (or ~8%) were associated with purely metric type II bursts and 16 (or ~11%) were associated with purely kilometric type II bursts. The average speeds CMEs associated with purely metric and purely km type II bursts are 740 km/s and 729 km/s, respectively. These are significantly below the average CME speed of all RL shocks (1237 km/s), but well above that of the RQ shocks (535 km/s). This means the RQ CMEs were not able to drive shocks that are strong enough to accelerate electrons even to produce purely m or purely km type II bursts. The CME speeds of the purely metric and kilometric type II bursts are consistent with the hierarchical relationship between CME kinetic energy and the wavelength range over which the type II bursts occur (Gopalswamy *et al* 2005). In the CME speed distribution of RL shocks, we have shown the average speed (1521 km/s) of CMEs faster than 900 km/s, which is consistent with other studies that considered fast and wide CMEs irrespective of IP shock association (Gopalswamy *et al* 2008b).

Table 3 summarizes the CME and shock speeds for all, RQ, and RL shocks divided according to the associated IP drivers (MC, EJ, or DL (Driverless)). We find that the MC-associated shocks have the lowest CME speed (891 km/s), the EJ associated shocks have intermediate CME speed (941 km/s), and the DL shocks have the highest CME speed (1308 km/s) (see table 3). This pattern applies not only to all the shocks, but also to the RQ and RL subsets, as can be seen in table 3. The differences in CME speeds can be attributed to the fact that the MC, EJ, and DL shocks have different solar source distributions: the MC-associated CMEs mostly originate close to the disk center, the EJ-associated CMEs originate at intermediate central meridian distances, and the CMEs responsible for the DL shocks originate close to the limb. Accordingly, the CMEs are subject to projection effects of varying extents: large (MC), intermediate (EJ), and minimal (DL). CMEs associated with the DL shocks are also expected be very energetic because they need to produce a shock signal at Earth in spite of their origin near the solar limb (see Gopalswamy *et al* 2007; 2009c). The shock speeds at 1 AU have a behavior opposite to that of the CME speeds: the MC-associated shocks are measured at their noses, so they are of the highest speed (625 km/s); the EJ-associated shocks are measured between their nose and flanks, so they are of intermediate speed (549 km/s); the DL shocks appear slow (average speed ~ 483 km/s) because they are actually the flanks of shocks propagating orthogonal to the Sun-Earth line. This pattern is also seen for RL and RQ shocks, with a single exception that the MC- and ejecta-associated RQ shocks have roughly the same speed.

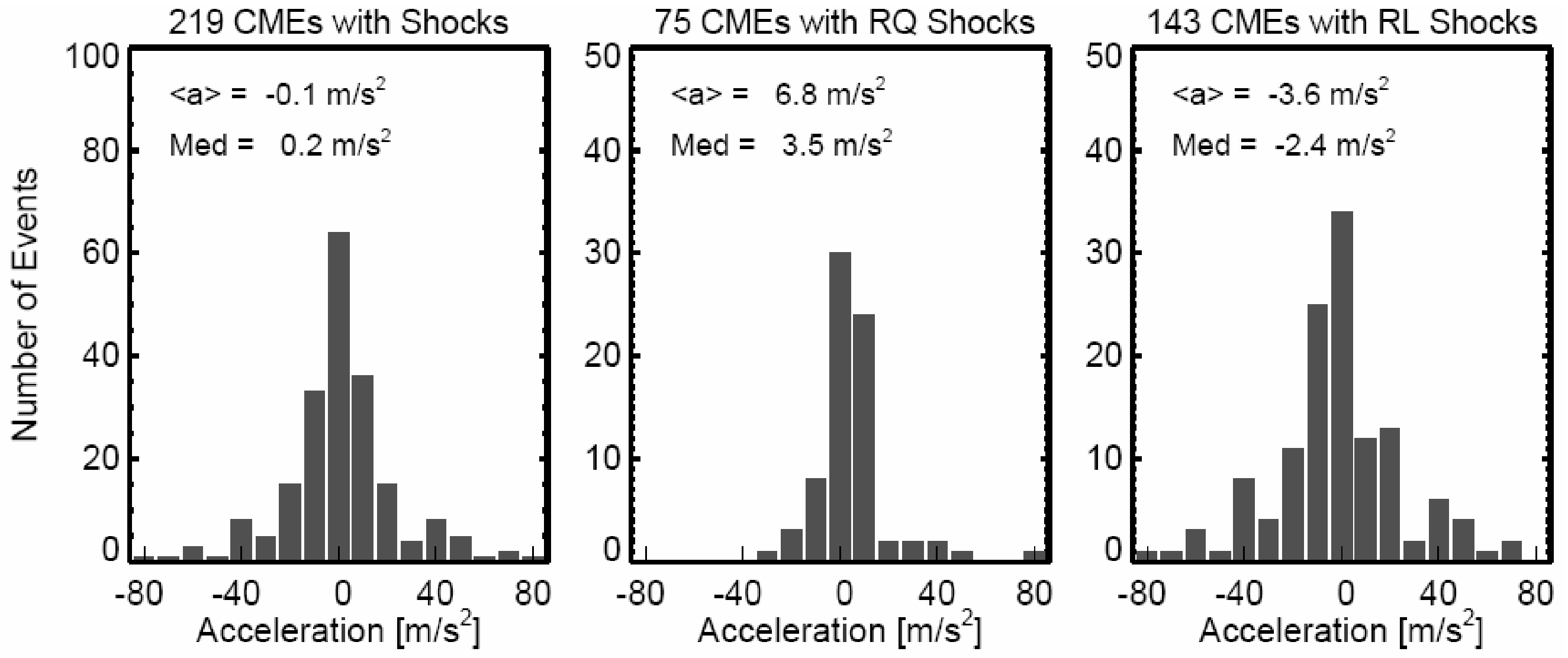

Figure 5. Distribution of CME accelerations for all shocks (left), RQ shocks (middle) and RL shocks (right). The average (<a>) and median (Med) values of the distributions are indicated on the plots. The number of events for which we were able to obtain the acceleration is also given on the plots. The bin size is 10 $ms^{-2}$ (e.g., the 40 $ms^{-2}$ bin contains events with acceleration from 35 $ms^{-2}$ to 45 $ms^{-2}$.

*4.2 CME width distributions*

Apart from the speed, the CME width is also an indicator of the CME energy. Halo CMEs are known to be faster on the average (Gopalswamy *et al* 2007), although we do not know their true width. However, we know that there is a good correlation between CME speed and width when CMEs erupting near the limb are considered (Gopalswamy *et al* 2009c). Therefore, we expect the halo CMEs to be wide and hence more energetic. Furthermore, CME width and mass are correlated (Gopalswamy *et al* 2005), so faster and wider CMEs generally have higher kinetic energy. Thus, the fraction of halo and partial halo CMEs in a population is a good indicator of how energetic the population is. The width distributions in figure 4 show that the fraction of full-halo CMEs is the largest for RL shocks (~72%) and the smallest for the RQ shocks (40%), differing by a factor of ~2. If we combine full (width = 360$^o$) and partial halos (width ≥120$^o$, but <360$^o$), we see that the fraction of such wide CMEs is still higher for the RL shocks, but to a smaller extent (97% vs. 72%). Only a small fraction of shocks are associated with non-halo CMEs, but even these CMEs are very wide: the average

CME width is 102° for RL shocks vs. 78° for RQ shocks. The speed and width distributions thus indicate that the CMEs driving the RQ shocks are definitely more energetic than the average CMEs, but less energetic than those driving RL shocks.

*4.3 CME acceleration distributions*

One of the striking differences in the kinematic properties of CMEs associated with the RQ and RL shocks is the sign of the acceleration within the coronagraphic field of view. Figure 5 shows that the average CME acceleration is close to zero negative (mean ~ - 0.1 $ms^{-2}$; median ~ 0.2 $ms^{-2}$) for all the shocks taken together. However, the average CME acceleration is positive for RQ shocks (mean ~ +6.8 $ms^{-2}$ and median ~ +3.5 $ms^{-2}$), while it is negative for RL shocks (mean ~ -3.6 $ms^{-2}$ and median ~ - 2.4 $ms^{-2}$). This result provides an important clue for understanding their radioquietness: the CMEs continue to accelerate and attain super-Alfvenic speeds only at a distance where the CME speed is high enough and the Alfven speed is low enough to form a shock. Even there, they seem to be only marginally super-Alfvenic so the shocks are not strong enough to accelerate electrons. Where such CMEs might become super-Alfvenic is illustrated in figure 6.

The CME acceleration measured for the two RQ events in figure 6 is close to the average CME acceleration of RQ shocks. The variation of the Alfven speed, the CME speed, and the solar wind speed are shown as a function of the heliocentric distance. The Alfven speed profile was obtained using models of magnetic field and plasma density as in Gopalswamy *et al* (2001a). The 2003 August 14 CME at 20:06 UT originated from S10E02 and had an average speed of ~378 km/s within the LASCO field of view. A second order fit to the height-time measurements yields a constant acceleration of 4.4 $ms^{-2}$. From the second order fit, we have plotted the speed as a function of distance in figure 6a. In the inner corona, the CME had a speed of only about 300 km/s, which is below the local Alfven speed. No shock is expected until about 8 Rs because of the higher Alfven speed in the ambient medium. The CME attained a speed of ~550 km/s when it reached the edge of the LASCO field of view.

Beyond 8 Rs, the CME maybe able to drive a shock provided the CME speed exceeds the sum of the Alfven speed and the solar wind speed.

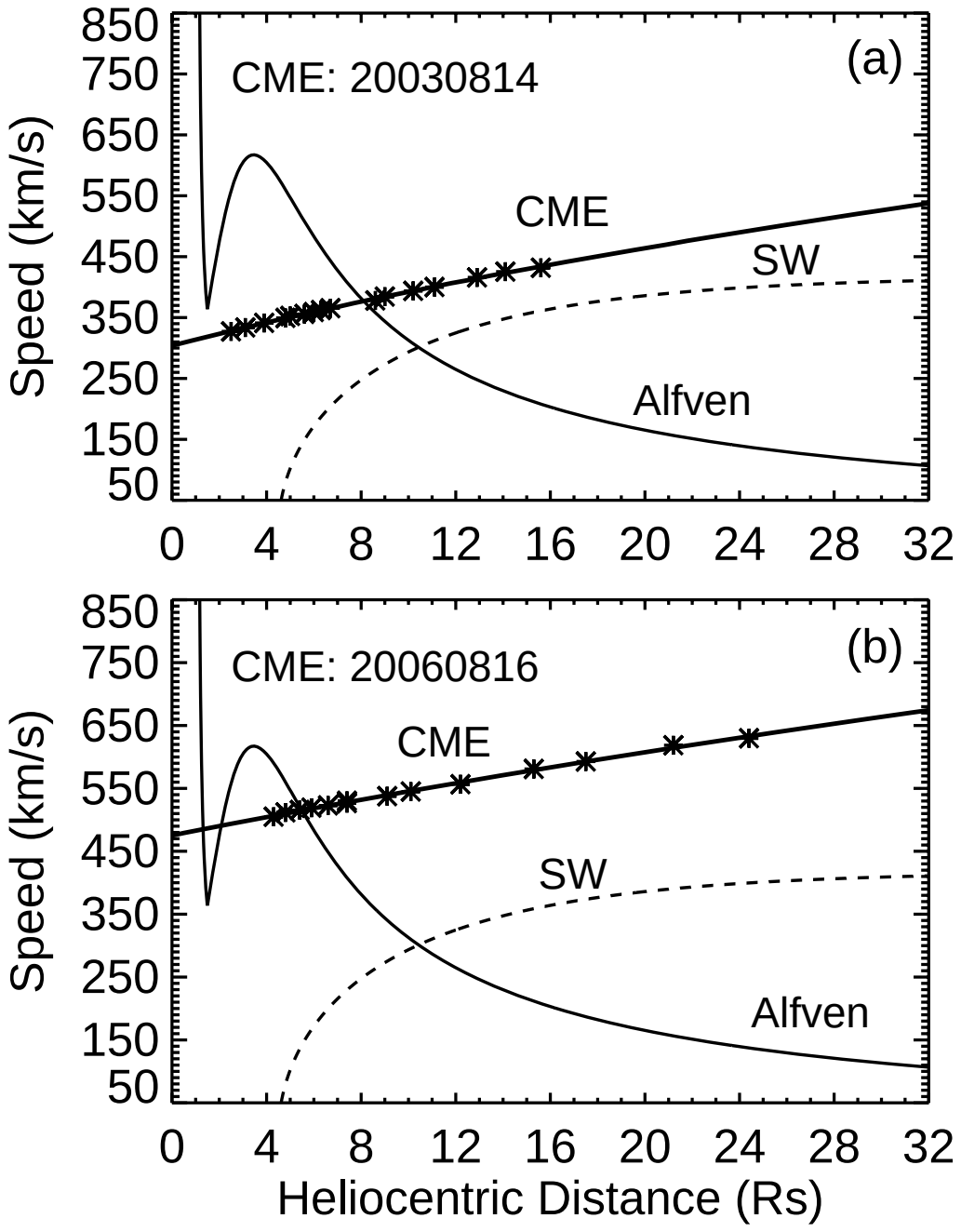

Figure 6. Variation of CME speed, Alfven speed, and solar wind (SW) speed as a function of heliocentric distance for (a) the 2003 August 14 CME at 20:06 UT, and (b) the 2006 August 16 CME at 07:31 UT. The data points are speed values from the second order fit at the heliocentric distances where height-time measurements were made.

The 2006 August 16 CME at 07:31 UT originated from the southwest quadrant (S01W19) and had an average speed of ~ 563 km/s within the LASCO field of view. A second order fit to the height-time measurements gives a constant acceleration of ~5.1 $ms^{-2}$. When plotted as a function of heliocentric distance (see figure 6b), the CME speed was found to be briefly above the Alfven speed in the inner corona, turning sub-Alfvenic due to higher Alfven speed until about 5.5 Rs, and then becomes super-Alfvenic. Once the CME speed exceeds the sum of the Alfven speed and the solar wind speed, a shock forms. Clearly the CME behavior (and its shock driving ability) is similar to the previous event beyond ~5.5 Rs. One would expect a metric type II burst in the 2006 August 16 event when the CME was super-Alfvenic briefly in the inner corona, but there was none. The Alfvenic Mach number (~1.3) was probably too small for the shock to be able to accelerate particles in sufficient numbers. It must be pointed out that the Alfven speed profiles shown in figure 6 were derived from models of density and magnetic field variation in the corona. The actual density and magnetic field on the days of the CMEs may be quite different. Both the CMEs in figure 6 also originated close to the disk center, so their true speeds may be higher. In this sense, the above discussion should be considered qualitative.

As noted before, 16 RL shocks had type II bursts only in the km wavelength domain. CMEs associated with 13 of these 16 shocks (or 81%) had positive acceleration, similar to those of the RQ shocks. These CMEs also must have attained super-Alfvenic speeds only far from the Sun, but their shocks are slightly stronger, so they produce type II emission. Interestingly, the subset of RL shocks that had radio emission only in the metric domain shows an opposite CME behavior: CMEs associated with 9 out of the 11 RL shocks (or 82%) had deceleration within the coronagraphic field of view suggesting that the shocks were efficient in electron acceleration only near the Sun (within the first 2-3 Rs) and remained subcritical thereafter.

The different acceleration profiles of RL and RQ CMEs obtained near the Sun, and the expected interaction with the solar wind can explain the reduced contrast between the RL and RQ shock properties at 1 AU (see figures 2 and 3). The evolution of CMEs during their interplanetary passage is such that fast CMEs tend to decelerate to the speed of the surrounding solar wind by the time they get to 1 AU (notable exceptions are the fastest events that have speeds well above the solar wind speed). On the other hand slow CMEs speed up to attain the solar wind speed. The two cases crudely correspond to the CMEs of RL and RQ shocks, respectively, consistent with the acceleration distributions in figure 5.

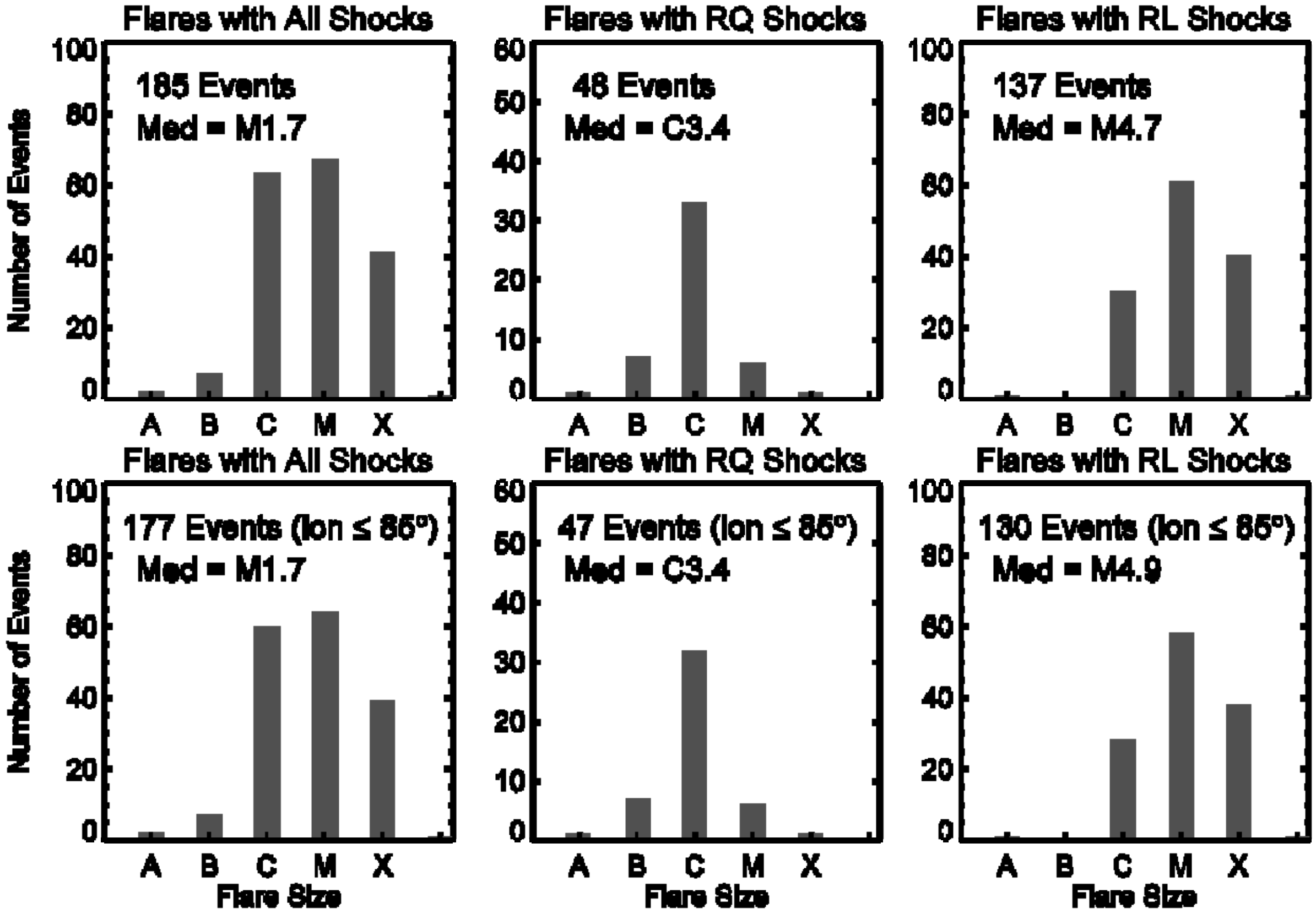

Figure 7. Size distribution of GOES soft X-ray flares associated with all (left), RQ (middle), and RL (right) shocks: (top) all flares for which position information is available and (bottom) flares occurring within 85$^o$ from the central meridian. The median flare sizes are shown on the plots. Note that the RQ and RL shocks have average flare sizes below and above that of all shocks. Each bin has a width of one order of magnitude: The bin C contains all C-class flares (i.e., flares with peak soft X-ray flux between $10^{-5}$ $Wm^{-2}$ and $10^{-4}$ $Wm^{-2}$).

*4.4 Flare sizes*

It was possible to identify the soft X-ray flares and their sizes (peak X-ray flux expressed as flare class) in the source regions of 48 RQ and 137 RL shocks. In several cases, the source locations were identified using the associated filament eruption, but the X-ray intensity was not above the background level, so no flare information is listed for these events. The flare-size distributions are shown in figure 7 for all shocks and the RQ and RL subsets. The median size of flares associated with RQ shocks (C3.4) is smaller by an order of magnitude compared to that in RL shocks (M4.7). The flares associated with RQ shocks are predominantly of C-class, while M- and X-class flares dominate in RL shocks (see figure 7). The difference in flare sizes is consistent with the difference in speeds and widths of RQ and RL CMEs (there is a reasonable correlation between X-ray flare size and CME kinetic energy – see Gopalswamy *et al* 2009c). We also examined the flare sizes for a subset of events in which the central meridian distance (CMD) of the flares ≤ 85$^o$ to avoid the possibility that some limb flares may be partially occulted and hence we may not know their true sizes. There were 47 such RQ shocks and 130 RL shocks whose median flare sizes are C3.4 and M4.9, respectively. These are not too different from the case when no CMD restriction is used.

*4.5 Solar sources of shocks*

The longitude and latitude distributions of the solar sources of the RQ and RL shocks are compared in figure 8. The numbers of RQ and RL shock sources decline towards limbs. The longitudinal distribution is mainly due to the selection effect because for a given CME speed, the shock from a disk-center CME has a better chance to be detected at 1 AU. For a disk-center eruption, the measurement is likely to be made at the nose of the shock compared to an eruption at a larger CMD. In the latter case, the measurement will be made at the shock flanks where the shock strength is typically lower (see discussion in section 4.1). The latitude distributions have a sharp cutoff around ±30$^o$ latitude for both subsets. Furthermore, the latitude distribution is bimodal, with peaks in the northern and southern active region belts. This suggests that the shock-driving CMEs originate in the active region belt. Active regions have a higher magnetic field, thus resulting in more energetic CMEs.

It must be pointed out that the longitude distribution of CME sources associated with RL shocks is quite different from the solar sources of RL CMEs reported in Gopalswamy *et al* (2008a). The center-to-limb variation of the number of fast and wide RL CMEs showed an increase, which is opposite to what is seen in figure 8. This can be explained by the fact that many limb and even behind-the-limb CMEs may produce type II bursts without producing a shock signal at Earth (Gopalswamy *et al* 2008a). On the other hand, the RQ and RL shocks have their solar sources clustered near the disk center (there are only six behind-the-limb events).

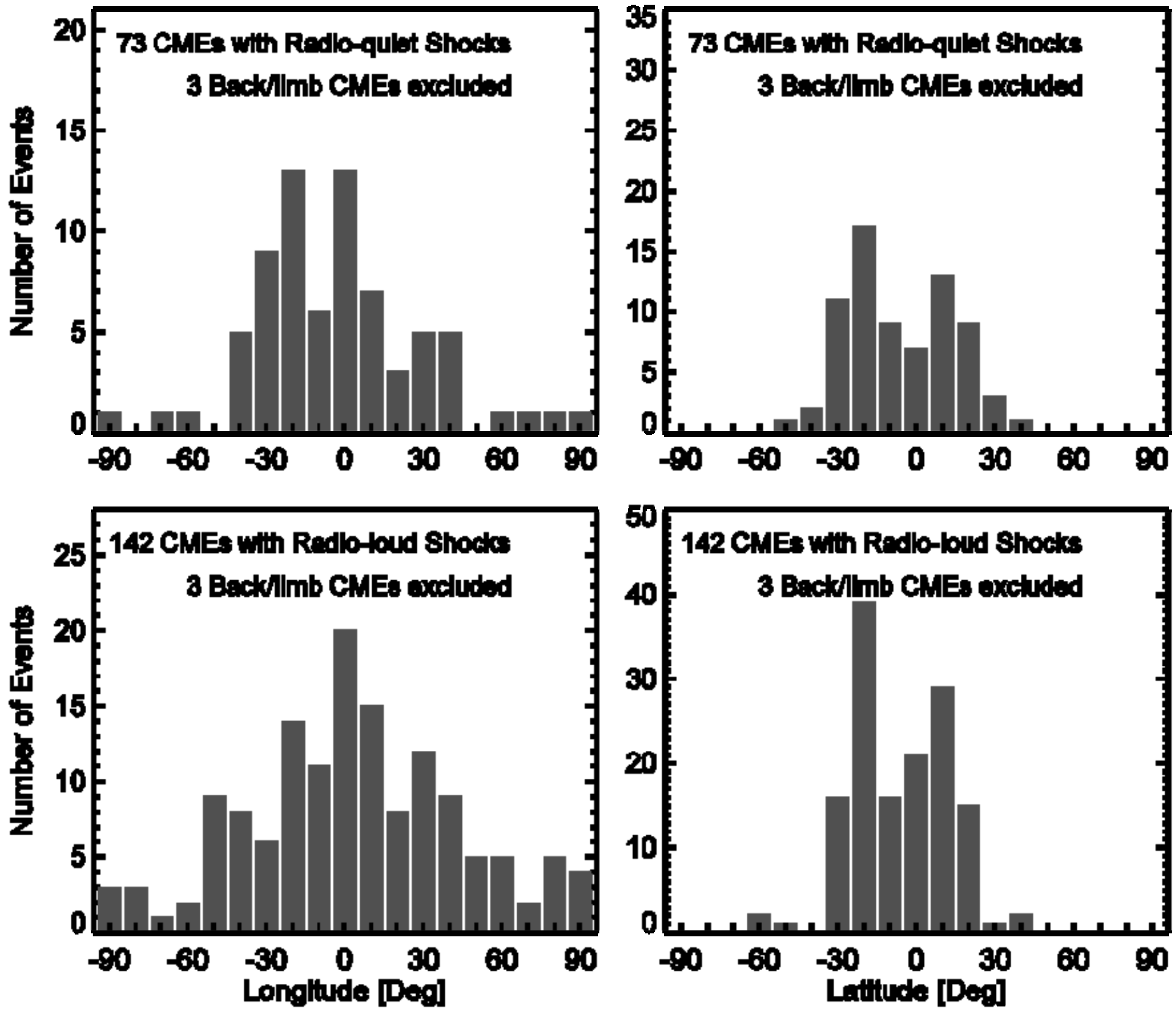

Figure 8. The longitude (left) and latitude (right) distributions of CME source regions for RQ (top) and RL (bottom) shocks. Three sources in each subset were behind the limb. The bin size is 10°. The 0° bin contains events from – 5° to +5°.

A different representation of the source distribution can be seen in figure 9, where the heliographic coordinates of the eruption regions are plotted for RQ and RL shocks separately and together. There are some clear differences between the source distributions: (i) there are almost no CME sources at the east limb for RQ shocks, whereas there are several RL sources at both limbs. The RQ sources are also more concentrated near the central meridian (with a slight eastern bias) compared to those of RL shocks, which show more uniform distribution in longitude (with a slight western bias). The limb sources of RL shocks are consistent with the fact that the associated CMEs are more energetic to produce a shock signature near Earth. (ii) The CME sources are generally within the ±30° latitude range, with only a small number of exceptions (the CME sources of 5 RQ shocks and 6 RL shocks outside - see also figure 8).

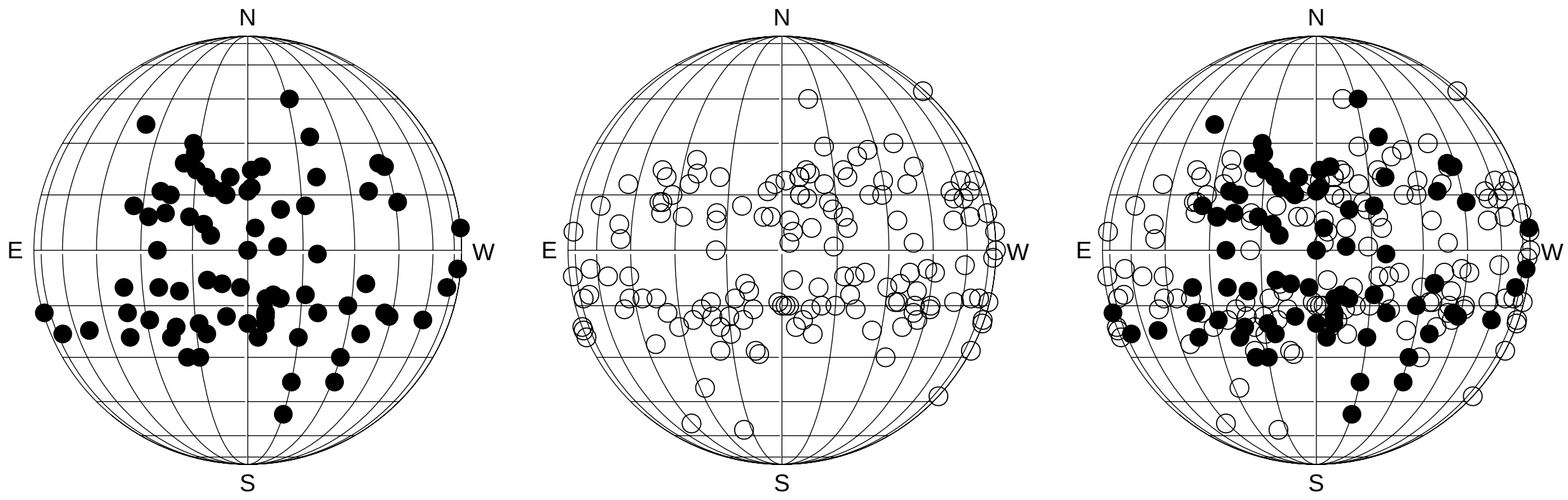

Figure 9. Heliographic coordinates of CMEs associated with RQ shocks (left) and RL shocks (middle) and the combined set (right).

*4.6 Solar-cycle variation of RQ and RL shocks*

The variation of source latitudes over the solar cycle shown in figure 10 explains the reason for the higher latitude sources in figure 9. The higher-latitude sources occur mostly during the rise phase of the solar cycle, similar to the solar sources of CMEs that produce a signature near Earth in the form of MCs or non-cloud ejecta (Gopalswamy *et al* 2008c; 2009d). This behavior has been attributed to the control by the global dipolar field, which has maximum strength during the solar minimum phase. In the beginning of a solar cycle, active regions emerge at higher latitudes, but the prominences and CMEs from these regions tend to move toward the equator under the influence of the global dipolar field (Gopalswamy *et al* 2000b; Gopalswamy and Thompson 2000; Filippov *et al* 2001; Plunkett *et al* 2001; Gopalswamy *et al* 2003a; Cremades *et al* 2006). During the rest of the solar cycle, the sources continue to follow the sunspot butterfly diagram.

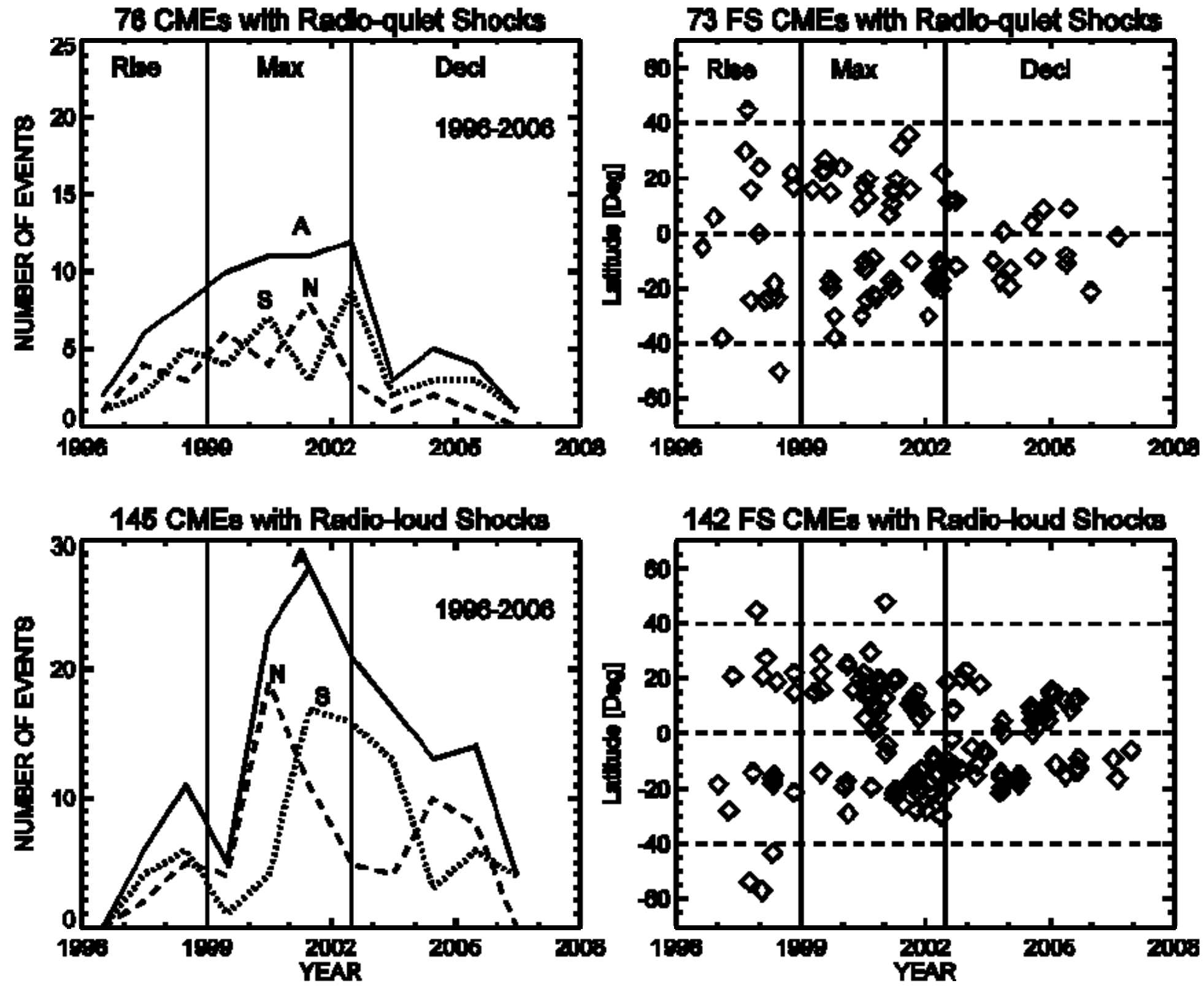

Figure 10. Variation of the rate of shock occurrence (left) and the source latitudes (right) as a function of time. Note that the latitude plots use only frontside (FS) events because we know their solar source locations. For behind the limb events included in the occurrence rate, we determined the hemisphere of occurrence from the position angle of the associated CMEs. The vertical solid lines delineate the different phases of the solar cycle (rise, maximum, and declining). The shock occurrence rate is given for all shocks (A) and for the northern (N) and southern (S) hemispheres.

Figure 10 shows that both RQ and RL shocks have a peak rate of occurrence during the maximum phase of the solar cycle (see also table 2). This simply reflects the fact that more energetic eruptions occur during the maximum phase. The active regions in which such eruptions occur are also closer to the equator. It has been shown that the average speed of CMEs increases by a factor of 2 between the minimum and maximum phases (see, e.g., Howard *et al* 1985; Gopalswamy 2004b; 2006a). The number of shocks in the northern and southern hemispheres also peak at different times. The latitude – time plot in figure 10 for RL shocks shows two clusters: one towards the end of year 2000 in the north and the other in the year 2002 in the south. High-latitude activities around these times are known to be associated with the solar polarity reversal (Gopalswamy *et al* 2003b), but it is not clear why such a high level of activity exists at low latitudes. One possibility is the presence of super active regions during these periods. On the other hand, the peaks in the RQ shock numbers are not very strong and seem to be due to statistical noise. Although the occurrence rates of RQ and RL shocks have a peak in the maximum phase, they have different behavior during the rise and declining phases: there are more RQ shocks during rise phase, contrary to the higher rate of RL shocks during declining phase. The number of RQ shocks drops sharply in the declining phase. This can be appreciated better when we plot the ratio of the number of RL shocks to that of the RQ shocks as a function of time (see figure 11). The ratio is ≤1 during the rise phase, increases to ~2 during the maximum phase and peaks at ~6 in the declining phase. There is a notable dip at the beginning of the maximum phase (year 1999), when there is a general dearth of energetic phenomena (see e.g., Gopalswamy *et al* 2004). The ratio starts decreasing in the late declining phase, but still remains above 1 towards the end of the study period. There are two possible explanations for the behavior of the RL to RQ shock number ratio. The first explanation is concerned with the solar cycle variation of Alfven speed. During the rise phase, the ambient medium typically has a lower density so the Alfvenic speed is expected to be higher, making it difficult to form strong shocks, thus rendering many shocks radio quiet. The 1-AU observation that there is no change in the Alfven speed between solar minima and maxima (Mullan and Smith 2006) poses problem to this explanation. However, we do not know if the solar cycle variation of the Alfven speed observed at 1 AU applies for the entire inner heliosphere where type II bursts are observed. The second explanation is the influence of super active regions, which are prolific producers of energetic CMEs and flares. During the declining phase of cycle 23, there were many super active regions that produced many energetic eruptions (Gopalswamy *et al* 2006). Shocks from these eruptions might have skewed their number in favor of RL shocks. It is not clear if such a behavior will be present in all cycles.

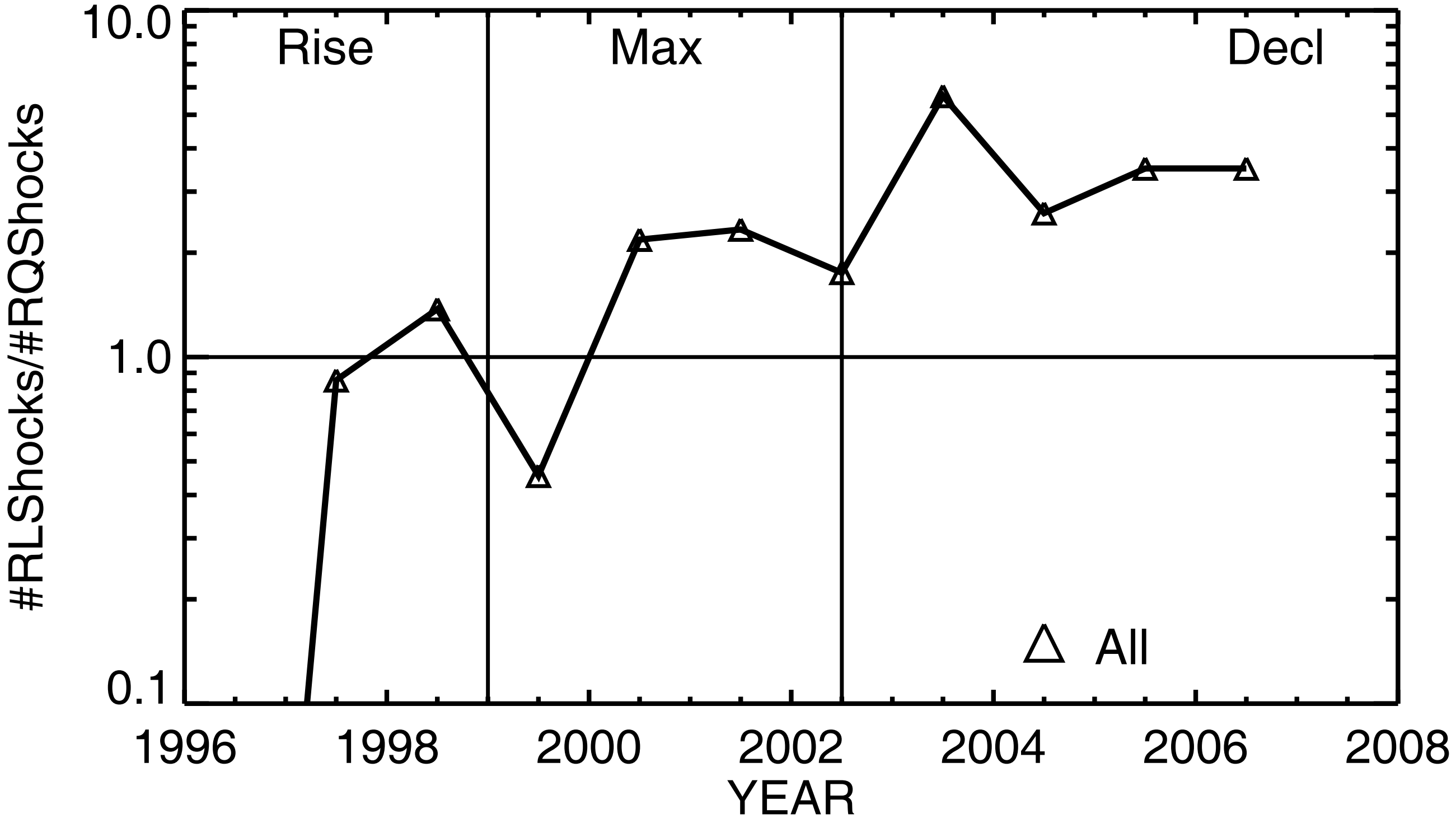

Figure 11. The ratio of the number of shocks in the RL subset to that in the RQ subset plotted as a function of time. The vertical lines delineate the phases of the solar cycle. The ratio was 0 (no RL shocks but three RQ shocks) during the year 1996, reached a maximum of 6 during 2003 and stayed above 1 until the end of the study period (2006).

**5. Discussion**

We investigated more than 200 interplanetary shocks detected by spacecraft near Earth and found that a large fraction (~34%) lacked type II emission in any of the wavelength domains. Type II bursts are the best indicators of CME-driven shocks propagating close to the Sun and in the IP medium. The CMEs associated with RQ shocks are of lower energy compared to those associated with the RL shocks. The two shock populations differ in many other aspects, as listed in table 4. As a general pattern, all the parameters of RQ CMEs point to a lower energy of the CMEs, resulting in weaker shocks and hence the lack of sufficient number of accelerated electrons needed to generate the type II radio bursts. However, the average speed of CMEs associated with RQ shocks exceeds that of the general population of CMEs, which means that the CMEs need to be of higher energy to be shock-driving. Furthermore, the fraction of full halos (40%) among the RQ CMEs is an order of magnitude larger than that in the general population (3.6%), thus confirming the higher energy of the shock-driving CMEs compared to the average CME.

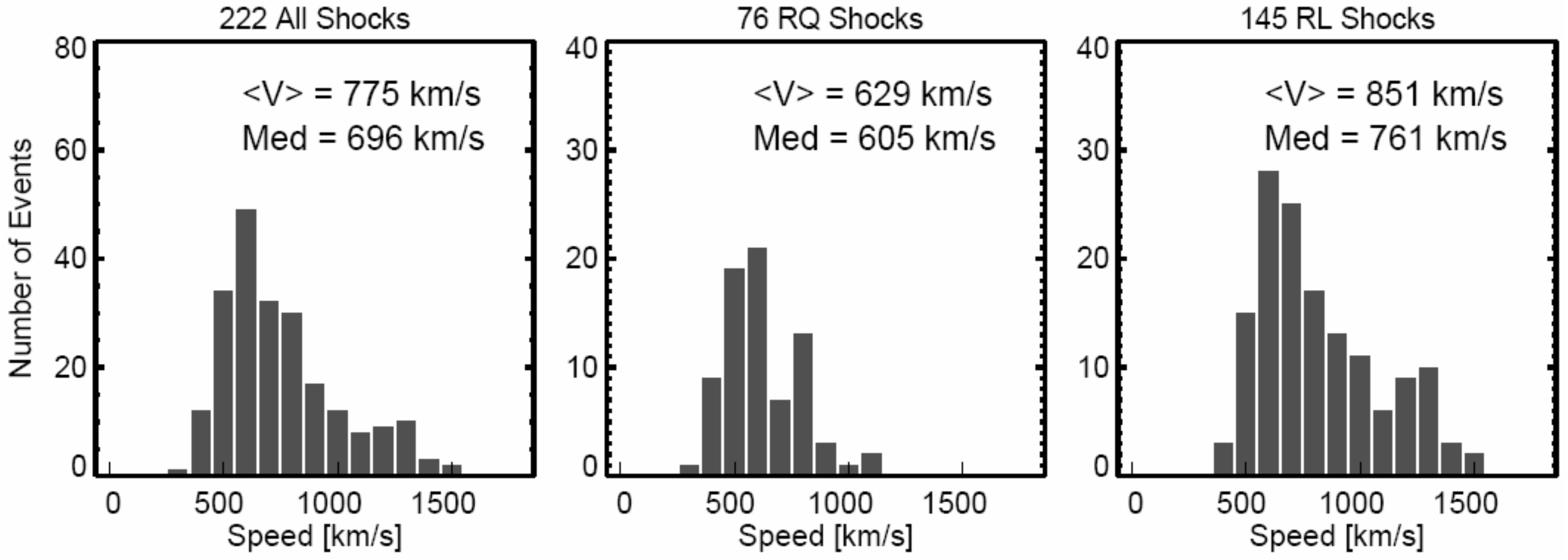

Figure 12. Distributions of shock transit speeds for all (left), RQ (middle) and RL (right) shocks. The average transit speeds (<V>) are smaller than the corresponding CME speeds for all shocks and RL shocks. However, the average transit speed for RQ shocks is less than that of the CMEs near the Sun, suggesting the peculiar kinematics of these CMEs. The bin size is 100 km/s. The 500 km/s bin contains shocks that have transit speeds in the range 450 to 550 km/s.

### *5.1 Shock transit speeds and CME kinematics*

In order to make a direct comparison with the past works on IP shocks, figure 12 shows the transit speed distributions of the shocks in our list. The average transit speed of all the shocks is ~775 km/s, with the RQ and RL shocks having the values 629 km/s and 851 km/s, respectively. The median transit speeds have similar relationships among the three populations. Note that these values are very similar to the average Sun-HELIOS transit speed (745 km/s) for a smaller number of shocks reported by Sheeley *et al* (1985) (see also Watari and Detman 1998 and references therein). The transit speeds are always greater than the in-situ shock speeds; they are also less than the corresponding CME speeds near the Sun, except for the RQ shocks. The RQ shocks thus point to a different kinematics as discussed below.

One of the distinct kinematic properties of the RQ shocks is the positive acceleration shown by most of the associated CMEs compared to the negative acceleration of the CMEs associated with RL shocks. This means that the shocks form generally at large distances from the Sun (≥10 Rs) where the driving CMEs become super-Alfvenic. CMEs associated with purely kilometric type II bursts show a similar kinematic behavior (Gopalswamy *et al* 2005; Gopalswamy 2006b). The km type II bursts typically start at frequencies below 1 MHz when the CMEs are typically at heliocentric distances >10 Rs. Thus the RQ shocks seem to be similar to those producing km type II bursts, except that the former are even weaker. One can see that the shocks are ordered according to the CME kinetic energy. The radio-quiet shocks are the weakest, followed by shocks producing the km type II bursts, and then the shocks producing radio emission at higher frequencies. There is also further hierarchy between CME kinetic energy and the frequency range of type II bursts reported elsewhere (Gopalswamy *et al* 2005). It is worth pointing out that the average speeds of CMEs associated with metric type II bursts (740 km/s) and km type II bursts (729 km/s) are similar, but the evolution of the underlying shocks is quite different: in the case of purely metric type II bursts the shock forms very close to the Sun and quickly becomes subcritical; in the case of km type II bursts, the shock forms far away from the Sun, similar to most of the RQ shocks. When Sheeley *et al* (1985) reported the connection between CMEs and IP shocks, the metric type II bursts were thought to be "neither necessary nor sufficient for the occurrence of interplanetary shocks". Here we see that purely metric type II bursts are indicative of shocks producing radio emission only near the Sun and remaining subcritical for the rest of the Sun-Earth distance. The IP shocks associated with purely metric type II bursts correspond to the lowest CME kinetic energy among all the RL shocks (see also Gopalswamy *et al* 2008c). When there is a metric type II with no associated shock at Earth, either the shock completely dissipates or it simply misses the spacecraft making in situ observations.

*5.2 Radio-quiet shocks vs. radio-quiet CMEs*

Radio-quiet CMEs have been studied before (Sheeley *et al* 1984; Gopalswamy *et al* 2001b; Lara *et al* 2003; Shanmugaraju *et al* 2003; Gopalswamy *et al* 2008b,c). Sheeley *et al* (1984) reported CMEs lacking metric type II bursts. Gopalswamy *et al* (2001b) reported that ~60% of CMEs with speeds exceeding 900 km/s were not associated with type II bursts in DH wavelengths, but the average CME width was relatively small. Lara *et al* (2003) considered CMEs associated with metric and DH type II bursts and confirmed that the speed and width of CMEs are important in deciding their type II association. Shanmugaraju *et al* (2003) confirmed the results of Gopalswamy *et al* (2001) and Lara *et al* (2003), using a sample of CMEs associated with flares of importance >C1.0. In addition, they point out the importance of flare duration, size, and rise time in the CME-type II association and contend that CMEs need not be the drivers of shocks producing type II bursts. Assuming that all type II bursts are due to CME-driven shocks, Gopalswamy *et al* (2008c) concluded that slow CMEs with type II busts and fast CMEs without type II bursts can be explained by the variation in the Alfven speed of the ambient medium. Thus, RQ CMEs either did not drive shocks, or the shocks were not able to accelerate electrons in sufficient numbers. In the case of RQ shocks, the underlying CMEs did drive shocks, so the inability to accelerate electrons seems to be the case. As noted before, many of the RQ shocks formed only at large distances (>10 Rs) from the Sun, so no shocks are expected in the corona.

*5.3 Alfvenic Mach numbers of RQ and RL shocks*

The low Mach number of the RQ shocks observed at 1 AU is consistent with the low-energy drivers associated with them. From numerical simulations, Burgess (2006) showed that supercritical shocks are required for accelerating electrons in the energy range relevant to the production of type II radio bursts. We suggest that the RQ shocks are subcritical, i.e., their Mach number is typically below the first critical Mach number (see Edmiston and Kennel 1984; Mann *et al* 2003): the critical Mach number ranges between 1 and 2 for quasi-parallel shocks and between 2.3 and 2.7 for quasi-perpendicular shocks for the coronal plasma. The actual value depends on the plasma beta of the ambient medium through which the shock propagates (Gary 2001). For example, the solar wind at 1 AU has beta ~1 (Mullan and Smith 2006), so the critical Mach number can be anywhere between 1 and 2.3 (Mann *et al* 2003), the higher values being for quasiperpendicular shocks. Recall that the average Mach number for RL shocks is 3.4, making them supercritical. On the other hand, the average Mach number for RQ shocks (2.6) is similar to the critical Mach number for quasi-perpendicular case, given the uncertainty in the estimation of Mach numbers (see section 3.1). Furthermore, the Mach numbers are estimated only at 1 AU in a high-beta situation. The Mach number in the near-Sun interplanetary medium and the corona can be very different because both the CME speed and ambient Alfven speed change with heliocentric distance. The density and magnetic field models of the corona indicate a peak Alfven speed of ~600 km/s at ~3 Rs from the Sun (see figure 6), which suggests an Alfven Mach number of ~6 for the fastest detected CMEs. For CMEs producing DH type II bursts, the average speed is ~1000 km/s indicating an Alfvenic Mach number of ~1.7 near the Sun. The corona contains dense streamers and tenuous regions, so the Alfven speed and Mach number can vary by a factor of 4 near the Sun (Gopalswamy *et al*, 2008c). Both the Alfven speed and the CME speed change (mostly decrease) with heliocentric distance, so does the Mach number. The existence of a type II burst thus depends on how these two speeds vary relative to each other between the Sun and Earth. The reduced contrast in Alfvenic Mach number between RQ and RL shocks can be attributed to the CME evolution in the inner heliosphere. It is well known that fast CMEs tend to decelerate to the speed of the surrounding solar wind by the time they get to 1 AU. On the other hand slow CMEs speed up to attain the solar wind speed. The two cases crudely correspond to the driving CMEs of RL and RQ shocks, respectively (the acceleration distributions are consistent with this).

### *5.4 RQ shocks and solar energetic particles*

It was previously shown that RQ CMEs were not associated with large SEP events, implying lack of ion acceleration. We expect something similar in the case of RQ shocks. Since the RQ shocks are likely to be subcritical, it is of interest to know whether energetic storm particle (ESP) events are also absent. Furthermore, RQ shocks seem to be quasiperpendicular, which can also be verified by determining the shock normal angle at 1 AU. A systematic investigation energetic particle data at the time of RQ shock arrival at 1 AU seem to support these conclusions, as reported in an accompanying paper (Mäkelä *et al* 2009, submitted). Such an analysis might explain the large fraction of IP shocks without ESP events reported by Ho *et al* (2008). The suggestion that most of the radio-quiet shocks may be quasi-perpendicular can be confirmed from the time profiles of ESP events, which substantially differ depending on the shock normal angle. In particular, quasi-perpendicular shocks are supposed to produce spike events (Sarris and van Allen 1974).

At the microscopic level, one expects a substantial difference between low and high Mach number shocks such as the mechanism of electron acceleration. In low Mach number shocks, the acceleration occurs via the classical fast Fermi acceleration, which increases the energy of the electrons only by a factor 2-3 and that too for nearly perpendicular shocks. High Mach number shocks acquire a dynamic rippled character and the energization can occur over a wider range of shock normal angles and to higher energy levels (see Burgess, 2006 for details). As noted before, we have very little information on the shock geometry except at the location of the observing spacecraft (at Sun-Earth L1 point). When the spacecraft passes through the shock, it might encounter superthermal particles (ESP events), which may provide information different from what the type II bursts provide. Type II bursts require electrons escaping from the shock front, while ESP events correspond to what is present at the shock front. Thus we expect some slightly different information on the shock strength from the ESP events and radio bursts. High Mach number shocks (supercritical) have also the possibility to accelerate electrons via the Bunemann instability of the reflected ion beam at the shock front (Cargill and Papadopoulos, 1988). Low Mach number (subcritical) shocks do not have ion reflection, so they cannot energize electrons by this mechanism. At 1 AU, the difference between the Mach numbers of RQ and RL shocks exists, but is not substantial. There are indications from CME observations that the difference may be substantial near the Sun.

The electron acceleration efficiency depends on additional factors also: the availability of seed particles and enhanced turbulence in the ambient medium into which the shock propagates.

*5.5 Interpretation of driverless shocks*

In a study aimed at identifying the IP drivers of shocks, Oh *et al* (2007) reported a small fraction of shocks in which they could not identify the IP driver. They discussed one of the "unidentified" driver events (2000 February 11 at 02:33 UT) in detail and suggested that it is a "blast-wave-type IP shock". Our investigation revealed that this shock is indeed associated with an IP ejecta, which is ~4 hours in duration (see table 1 and figure 13). The ejecta is clearly identifiable from the temperature signature as shown in figure 13. There was also a second shock on this day associated with another ejecta. The solar sources of the two CMEs are in the NE and SW quadrants, respectively on 2000 February 8 at 09:30 UT and on the next day at 19:54 UT as shown in figure 13. The two CMEs were also associated with IP type II bursts, and hence were driving shocks near the Sun. Therefore, we conclude that the blast-wave scenario is incorrect in this event because the IP shock has a driver identified both in the IP medium and near the Sun.

In all, Oh *et al* (2007) reported that 10 IP shocks had "unidentified" drivers. Five of these shocks are not in our list, probably because they are either CIR shocks or slow shocks. Of the remaining five, two are in our list as driverless shocks (2007 October 23 and 2001 March 27 shocks for which we know the CME driver at the Sun); one was associated with an MC (2000 February 11 shock) and two were associated with ejecta (1998 May 3 and 2001 September 13 shocks). Note that we were able to identify the driving CMEs near the Sun for all the 42 IP shocks with no discernible IP drivers. The reason for the lack of IP drivers is simply the large CMD of the eruption regions, such that we just observe the shock flanks at Earth. The drivers head roughly orthogonal to the Sun-Earth line, so they are not intercepted by the observing spacecraft (see Gopalswamy *et al* 2001a; 2009a). Thus none of the driverless shocks qualifies to be a blast wave.

During our study period overlapping with that of Oh *et al* (2007), we identified 14 driverless shocks. Four of these did not figure in Oh *et al* (2007) list: (1996 August 16 and November 11, 2000 July 26, and 2001 April 7). As we noted above, only two were designated as "unidentified" by Oh *et al*. They identified six of the shocks with MCs, one with a high speed stream and one with an ejecta. Thus there is clearly disagreement on the IP drivers of at least 8 shocks. Our driverless shock on 2001 January 31 was designated as a HSS shock by Oh *et al* (2007), which we think is incorrect. The speed never increased more than 450 km/s for the next two days. If anything, there was a tiny ejecta (1-2 h long) around 18 UT on 2001 February 1, which is more than 34 h after the shock. Even if this ejecta is associated with the shock, it is consistent with the driverless shock classification (the driver is barely seen). The solar source was a CME on 2001 January 28 at 15:54 UT from S04W59, which suggests that just the eastern flank of the shock should have arrived at Earth. In addition to these differences, we note a predominance of MCs as drivers in Oh *et al* (2007) list. This may be because they seem to have combined both MC and cloud-like events as MCs. We designated the cloud-like events as ejecta according to the list maintained by R. P. Lepping (http://lepmfi.gsfc.nasa.gov/mfi/MCL1.html).

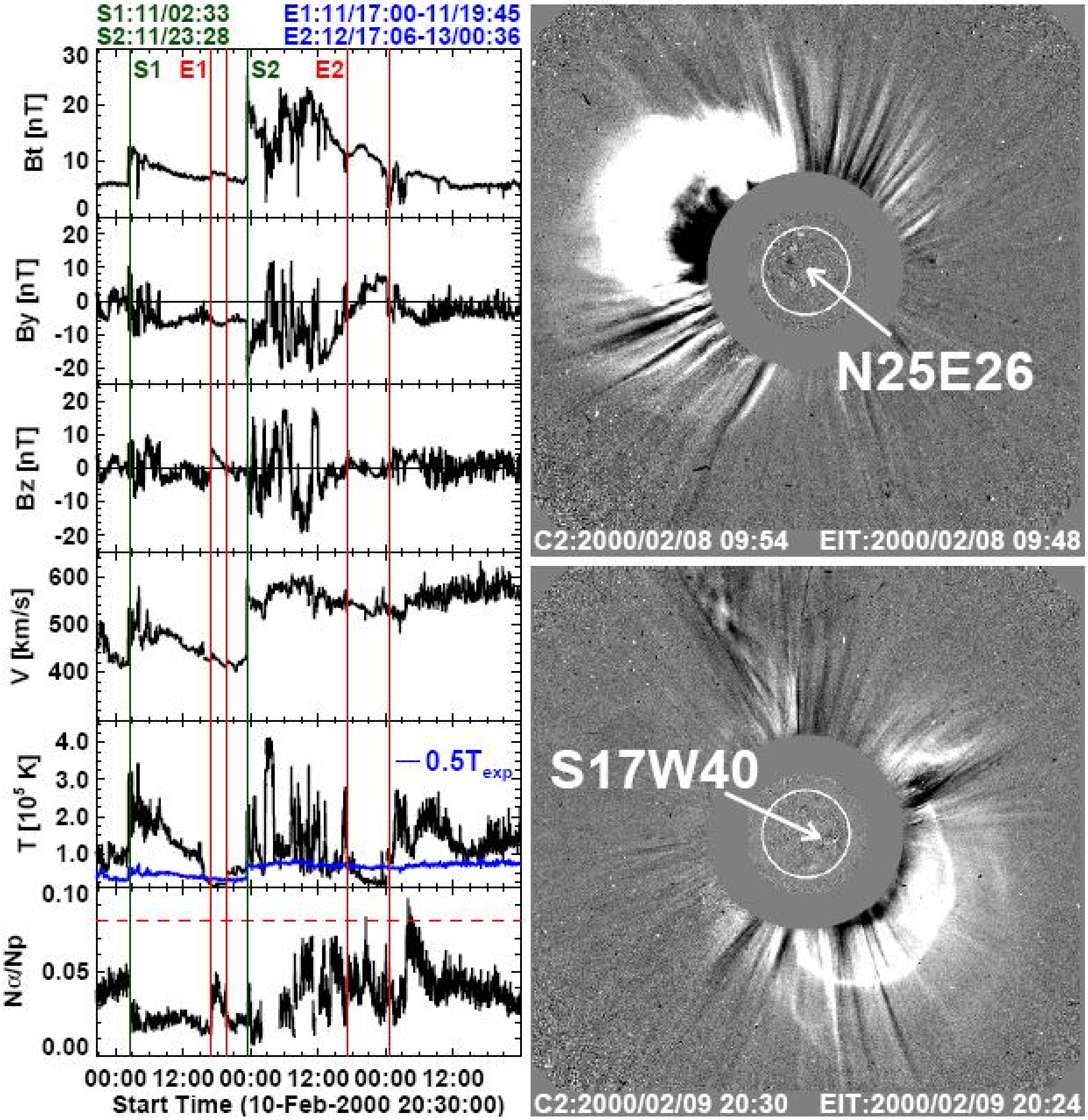

Figure 13. (left) IP shocks (S1, S2) on 2000 February 11 at 02:33 and 23:28 UT with the corresponding ejecta (E1, E2). The ejecta intervals are identified as those for which the proton temperature drops below 0.5 Texp. The shock times and ejecta intervals are marked by vertical lines. (right) Snapshots of CMEs associated with the IP shocks. The source regions of the CMEs are indicated by the arrows.

## 6. Summary and conclusions

We compiled and analyzed the solar-source properties (CMEs, flares, and radio-burst association) of 222 interplanetary shocks and found that a large fraction of them (~34%) was radio quiet (lacked type II radio bursts). We arrived at the following conclusions when we compared the RQ shocks with the RL ones. (1) The primary characteristic that distinguishes RQ and RL shocks seems to be the kinetic energy of the CME drivers. The lower CME kinetic energy of RQ shocks is also suggested by the lower peak soft X-ray flux of the associated flares. (2) CMEs associated with the RQ shocks have different average kinematic properties near the Sun: they show positive acceleration, in contrast to the CMEs associated with RL shocks. This means the underlying CMEs become super-Alfvenic and drive shocks only at large distances from the Sun (typically >10 Rs). (3) Based on the CME acceleration, we see that the RQ shocks seem to be similar to the RL shocks producing kilometric type II bursts, but weaker. (4) Among the RL shocks, those producing purely metric and purely kilometric type II bursts have the lowest CME kinetic energy, but have opposite kinematic evolution: CMEs associated with metric type II bursts seem to be accelerating electrons near the Sun, while those associated with kilometric type II bursts accelerate electrons far away from the Sun. (5) Although we do not know the shock Mach numbers near the Sun, the 1-AU values indicate that the RQ shocks may be subcritical because of their inability to accelerate electrons to ~0.2 - 10 keV needed to produce type II bursts. The critical Mach number is typically <2.7, so the average Mach number 2.6 makes the RQ shocks subcritical. On the other hand, the RL shocks are supercritical with an average Mach number of 3.4. Given the uncertainty in the determination of Mach numbers, the difference is modest. The contrast between RQ and RL shocks seems to be erased during the interplanetary propagation of CMEs with different kinematics. (6) The average Mach number of RQ shocks is near the higher end of the typical range for critical Mach numbers suggesting that the RQ shocks may be predominantly quasi-perpendicular at Earth. (7) There is a clear solar-cycle variation of radio-quietness: there are more RQ CMEs in the rise phase compared to those in the maximum and declining phases. It is not clear if this property is peculiar to solar cycle 23 because of the super active regions or it is due to the change in the physical conditions in the interplanetary medium over the solar cycle. (8) About 19% of the IP shocks in our list were “driverless”, i.e., these shocks were not followed by discernible ICMEs at Earth. All shocks were associated with CMEs that were ejected at large central meridian distances, except for a few cases, which were deflected by coronal-hole magnetic fields. Therefore, we can rule out the possibility that the “driverless” shocks are blast waves. (9) The solar sources of the shock-driving CMEs follow the sunspot butterfly diagram, consistent with the higher-energy requirement for driving shocks.

**Acknowledgements:** We thank J. C. Kasper and A. Viñas for their help in obtaining shock Mach numbers. This work was supported by NASA’s LWS TR&T and SR&T programs. We thank the anonymous referee for critical comments that helped improve the presentation of the paper.

Table 1. List of IP shocks and the associated phenomena

[The complete version of this table is in the electronic edition of the Journal. The printed edition contains only a sample.]

Table 1 (Electronic Supplement): List of IP shocks observed during solar cycle 23 and their source information

| IP SHOCK | | | | | | ICME | | | | CME | | | | | | Solar Flare | | | Type II Burst | | | | Notes |
|---|---|---|---|---|---|---|---|---|---|---|---|---|---|---|---|---|---|---|---|---|---|---|---|
| | | | | | | | | | | | | | | | | | | | Metric | | DH | | |
| Date | Time | S/C | Speed | Ma | C | Type | Tstd | Time | Speed | Date | Time | CPA | W | Speed | a | Loc | Time | Size | C | Time | C | Time | |
| 19960816 | 07:45 | W | 390.50 | 3.44 | 1 | S | ---- | ----- | --- | 19960814 | 19:30 | 229 | 235 | 691 | --- | S05W80 | 1820-1907 | A3.0 | 0 | ---- | 0 | ---- | |
| 19961111 | 15:12 | W | 358.20 | 3.43 | 1 | S | ---- | ----- | --- | 19961107 | 23:20 | Halo | 360 | 497 | 8.7 | SElimb | 2252----- | DIM | 0 | ---- | 0 | ---- | |
| 19961202 | 10:00 | W | 329.30 | 5.23 | 1 | S | ---- | ----- | --- | 19961128 | 16:50 | 267 | 101 | 984 | 27.5 | N06W90 | 1535-1902 | C1.3 | 0 | ---- | 0 | ---- | |
| 19970110 | 00:52 | W,S | 433.70 | 5.40 | 1 | MC | 4.4 | 05:18 | 436 | 19970106 | 15:10 | Halo | 360 | 136 | 4.1 | S18E06 | 1454-1542 | A1.1 | 0 | ---- | 2 | 0200 | |
| 19970209 | 12:40 | W,S | 617.60 | 2.93 | 1 | MC | 14.7 | *03:24 | 458 | 19970207 | 00:30 | Halo | 360 | 490 | 14.3 | S38W31 | 2306----- | EP | 0 | ---- | 0 | ---- | |
| 19970410 | 17:57 | W,S | 411.23 | 1.86 | 1 | MC | 11.7 | *05:36 | 460 | 19970407 | 14:27 | Halo | 360 | 878 | 3.3 | S28E19 | 1350-1419 | C6.8 | 1 | 1358 | 3 | 1430 | 1 |
| 19970515 | 01:15 | W,S | 443.20 | 4.05 | 1 | MC | 7.8 | 09:06 | 450 | 19970512 | 05:30 | Halo | 360 | 464 | -15.0 | N21W08 | 0442-0526 | C1.3 | 1 | 0454 | 3 | 0515 | |
| 19970902 | 22:40 | W,A,S | 368.00 | 3.49 | 1 | EJ | 13.3 | *12:00 | 400 | 19970830 | 01:30 | Halo | 360 | 371 | 9.3 | N30E17 | 2256-2354 | M1.4 | 0 | ---- | 0 | ---- | |
| 19970921 | 04:10 | W,A | 273.62 | 1.15 | 2 | MC | 20.6 | *00:48 | 425 | 19970917 | 20:28 | Halo | 360 | 377 | 0.0 | N45W16 | 1802----- | EP | 0 | ---- | 0 | ---- | 2 |
| 19971010 | 16:00 | W,A | 470.70 | 1.61 | 1 | MC | 7.8 | 23:48 | 396 | 19971006 | 15:28 | 139 | 174 | 293 | 15.9 | S54E46 | 1332----- | EP | 0 | ---- | 2 | 0520 | 3 |
| 19971023 | 08:10 | W,A,S | 386.39 | 1.60 | 2 | S | ---- | ----- | --- | 19971021 | 01:27 | 108 | 91 | 570 | 0.9 | S24E37 | 0050----- | EP | 0 | ---- | 0 | ---- | |
| 19971024 | 11:15 | W,A,S | 489.80 | 2.14 | 1 | EJ | 5.8 | 17:00 | 420 | 19971021 | 18:03 | Halo | 360 | 523 | -2.9 | N16E07 | 1700-1816 | C3.3 | 0 | ---- | 0 | ---- | |
| 19971106 | 22:10 | W,A,S | 402.63 | 4.05 | 2 | MC | 17.6 | *15:48 | 440 | 19971104 | 06:10 | Halo | 360 | 785 | -22.1 | S14W33 | 0552-0602 | X2.1 | 1 | 0558 | 3 | 0600 | |
| 19971122 | 09:10 | W,A,S | 489.40 | 4.57 | 1 | MC | 6.6 | 15:48 | 490 | 19971119 | 17:00 | DG | DG | DG | DG | N20E05 | 1700-2140 | C1.6 | 0 | ---- | 0 | ---- | |
| 19971210 | 04:30 | W,S | 374.70 | 2.81 | 1 | EJ | 14.2 | 18:40 | 375 | 19971206 | 10:27 | 317 | 223 | 397 | 9.0 | N45W10 | 1000----- | EP | 0 | ---- | 2 | 0730 | 3 |
| 19971230 | 01:15 | W,A,S | 385.70 | 2.74 | 1 | EJ | 16.8 | 18:00 | 370 | 19971226 | 02:31 | 186 | 230 | 197 | 5.5 | N00E25 | --------- | --- | 0 | ---- | 0 | ---- | 4 |
| 19980106 | 13:30 | W,A,S | 406.10 | 2.75 | 1 | MC | 13.8 | *03:18 | 375 | 19980102 | 23:28 | Halo | 360 | 438 | 6.5 | N24W42 | 2335-0630 | B6.4 | 0 | ---- | 0 | ---- | |
| 19980124 | 04:40 | W,A,S | 414.20 | 4.08 | 1 | EJ | 1.0 | 05:40 | 400 | 19980121 | 06:37 | Halo | 360 | 361 | 1.5 | S57E19 | 0419----- | EP | 0 | ---- | 2 | 1040 | 5 |
| 19980128 | 16:00 | W,A,S | 353.18 | 2.98 | 2 | EJ | 21.8 | *13:45 | 390 | 19980125 | 15:26 | Halo | 360 | 693 | -7.4 | N21E25 | 1429-1700 | C1.1 | 0 | ---- | 1 | 1503 | |
| 19980218 | 07:50 | W,A,S | 450.70 | 1.19 | 1 | EJ | 13.3 | 21:10 | 370 | 19980214 | 06:55 | 137 | 206 | 123 | 0.7 | S24E23 | 0326-0341 | B2.8 | 0 | ---- | 0 | ---- | |
| 19980304 | 11:05 | W,A | 433.10 | 2.94 | 1 | MC | 3.2 | 14:18 | 360 | 19980228 | 12:48 | 236 | 169 | 176 | 2.9 | N28W27 | 0905----- | EP | 0 | ---- | 2 | 0630 | 3 |
| 19980407 | 16:55 | W,A,S | 368.80 | 2.04 | 1 | EJ | 7.1 | *00:00 | 330 | 19980404 | 02:46 | 18 | 58 | 237 | 1.1 | S23E12 | 2335-0002 | C6.7 | 0 | ---- | 0 | ---- | 5,6 |
| 19980423 | 17:30 | W,A,S | 401.50 | 3.26 | 1 | S | ---- | ----- | --- | 19980420 | 10:07 | 284 | 165 | 1863 | 43.5 | S43W90 | 0938-1118 | M1.4 | 1 | 0956 | 3 | 1025 | |
| 19980430 | 08:40 | W,A | 307.50 | 15.74 | 1 | S | ---- | ----- | --- | 19980427 | 08:56 | Halo | 360 | 1385 | 74.4 | S16E50 | 0855-0938 | X1.0 | 1 | 0908 | 3 | 0925 | |
| 19980501 | 21:20 | W,A,S | 630.70 | 2.48 | 1 | MC | 15.0 | *12:18 | 515 | 19980429 | 16:58 | Halo | 360 | 1374 | -44.8 | S18E20 | 1606-1659 | M6.8 | 1 | 1622 | 1 | 1630 | |
| 19980503 | 17:00 | W,A,S | 473.90 | 3.39 | 1 | EJ | 2.0 | 19:00 | 480 | 19980501 | 23:40 | Halo | 360 | 585 | 8.0 | S18W05 | 2236-2308 | M1.2 | 0 | ---- | 0 | ---- | 7 |
| 19980504 | 02:00 | W,A,S | 356.95 | 2.17 | 2 | EJ | 10.0 | 12:00 | 650 | 19980502 | 14:06 | Halo | 360 | 938 | -28.8 | S15W15 | 1331-1351 | X1.1 | 1 | 1341 | 1 | 1425 | |
| 19980515 | 13:50 | W,A,S | 320.20 | 5.08 | 1 | S | ---- | ----- | --- | 19980512 | 08:55 | 210 | 225 | 1073 | -8.3 | SWlimb | 0812----- | DIM | 0 | ---- | 0 | ---- | 6,8 |
| 19980529 | 15:10 | W,A,S | 694.90 | 2.22 | 1 | EJ | 7.6 | 22:46 | 650 | 19980527 | 13:45 | 175 | 268 | 878 | -3.7 | N19W62 | 1330-1450 | C7.5 | 0 | ---- | 1 | 1330 | |
| 19980613 | 19:20 | W,A,S | 267.50 | 2.07 | 1 | EJ | 2.7 | *02:00 | 360 | 19980609 | 09:27 | 201 | 130 | 124 | 2.7 | S23W35 | 0835----- | EP | 0 | ---- | 0 | ---- | |
| 19980625 | 16:10 | W,A | 352.50 | 1.19 | 1 | EJ | 9.8 | *02:00 | 470 | 19980622 | 07:34 | 191 | 119 | 278 | 6.7 | S50W15 | 0306----- | EP | 0 | ---- | 0 | ---- | |
| 19981018 | 19:28 | W,A | 321.80 | 3.92 | 1 | MC | 9.6 | *05:06 | 383 | 19981015 | 10:04 | Halo | 360 | 262 | 3.2 | N22W01 | 0211----- | EP | 0 | ---- | 0 | ---- | |
| 19981107 | 08:00 | W,A,S | 508.80 | 2.07 | 1 | EJ | 13.9 | 21:55 | 480 | 19981104 | 07:54 | Halo | 360 | 523 | 19.6 | N17W01 | 0713-0731 | C1.6 | 0 | ---- | 0 | ---- | |
| 19981108 | 04:42 | W,A,S | 644.70 | 1.52 | 1 | MC | 19.1 | 23:48 | 470 | 19981105 | 20:44 | Halo | 360 | 1118 | -24.0 | N22W18 | 1900-2012 | M8.4 | 1 | 1950 | 3 | 2200 | |
| 19981112 | 07:18 | W,A | 406.10 | 1.29 | 1 | EJ | 13.5 | 20:48 | 350 | 19981108 | 11:54 | 264 | 196 | 559 | 6.2 | S21W37 | 1055-1131 | C5.9 | 0 | ---- | 1 | 1120 | |
| 19981113 | 01:40 | W | 395.86 | 1.14 | 2 | EJ | 2.8 | 04:25 | 410 | 19981109 | 18:18 | 330 | 190 | 325 | 2.6 | N15W05 | 1703-1852 | C2.5 | 0 | ---- | 2 | 0100 | 9 |
| 19990310 | 01:30 | W,A,S | 501.90 | 3.69 | 1 | EJ | 16.0 | 17:30 | 450 | 19990307 | 03:54 | ---- | --- | ---- | ---- | S24W03 | 0354-0415 | C2.9 | 0 | ---- | 0 | ---- | 10 |
| 19990416 | 11:10 | W,A,S | 455.50 | 1.65 | 1 | MC | 9.1 | 20:18 | 406 | 19990413 | 03:30 | 228 | 261 | 291 | 0.2 | N16E00 | 0145-0250 | B4.3 | 0 | ---- | 0 | ---- | |
| 19990505 | 14:59 | A,S | 406.30 | 2.72 | 1 | EJ | 13.0 | *04:00 | 400 | 19990503 | 06:06 | Halo | 360 | 1584 | 15.8 | N15E32 | 0536-0632 | M4.4 | 1 | 0550 | 3 | 0550 | |
| 19990626 | 02:30 | W,A,S | 371.20 | 3.42 | 1 | EJ | 3.5 | 06:00 | 350 | 19990622 | 18:54 | Halo | 360 | 1133 | -24.7 | N22E37 | 1737-1854 | M1.7 | 1 | 1824 | 3 | 1825 | |
| 19990626 | 19:25 | W,A,S | 467.60 | 2.19 | 1 | EJ | 11.9 | *07:20 | 470 | 19990624 | 13:31 | Halo | 360 | 975 | 32.4 | N29W13 | 1204-1510 | C4.1 | 0 | ---- | 1 | 1430 | |
| 19990702 | 00:23 | W,A,S | 633.00 | 2.42 | 1 | EJ | 5.6 | 06:00 | 600 | 19990629 | 19:54 | Halo | 360 | 560 | -8.9 | S14E01 | 1907-1920 | M1.6 | 0 | ---- | 1 | 1920 | |
| 19990706 | 14:24 | W,A,S | 473.60 | 1.71 | 1 | EJ | 7.1 | 21:30 | 450 | 19990703 | 19:54 | 301 | 139 | 536 | 3.1 | N16W55 | 2015-2059 | C7.5 | 0 | ---- | 3 | 1942 | |
| 19990712 | 01:20 | W,A,S | 217.60 | 1.75 | 2 | S | ---- | ----- | --- | 19990707 | 19:31 | 303 | 163 | 547 | 0.1 | N23W44 | 1736----- | EP | 0 | ---- | 0 | ---- | |
| 19990804 | 01:46 | W,A,S | 409.40 | 2.15 | 1 | EJ | 2.9 | 04:40 | 380 | 19990801 | 21:50 | 23 | 101 | 347 | 12.4 | N27E16 | 2059-2130 | C3.5 | 0 | ---- | 0 | ---- | |
| 19990808 | 17:44 | W,A | 437.00 | 1.13 | 1 | MC | 17.1 | *10:48 | 370 | 19990803 | 05:50 | 342 | 59 | 222 | 6.1 | N23W04 | 0435-0444 | C4.0 | 0 | ---- | 0 | ---- | 11 |
| 19990915 | 07:43 | W,A,S | 667.90 | 2.77 | 1 | EJ | 3.1 | 10:48 | 630 | 19990912 | 00:54 | 271 | 121 | 732 | 9.2 | S17W42 | 0012----- | EP | 0 | ---- | 0 | ---- | |
| 19990915 | 20:00 | W,A,S | 553.60 | 3.19 | 1 | EJ | 7.0 | *03:00 | 575 | 19990913 | 17:31 | 109 | 184 | 444 | -8.7 | N15E06 | 1630-1704 | C2.6 | 0 | ---- | 0 | ---- | |
| 19990922 | 12:00 | W,A | 465.90 | 2.68 | 1 | EJ | 9.2 | 21:14 | 600 | 19990920 | 06:06 | Halo | 360 | 604 | -14.5 | S20W05 | 0358----- | EP | 0 | ---- | 0 | ---- | |
| 19991021 | 02:13 | W,A,S | 476.00 | 2.52 | 1 | EJ | 7.4 | 09:35 | 410 | 19991018 | 00:06 | 206 | 240 | 144 | 3.5 | S30E15 | 2322-2336 | C1.2 | 0 | ---- | 0 | ---- | |
| 19991028 | 12:10 | W,A,S | 428.60 | 1.68 | 1 | EJ | 14.8 | *03:00 | 380 | 19991025 | 14:26 | 186 | 146 | 511 | 4.0 | S38W15 | 1313----- | EP | 0 | ---- | 0 | ---- | |
| 19991226 | 21:30 | A | 430.30 | 1.00 | 1 | EJ | 21.2 | *18:43 | 450 | 19991222 | 19:31 | Halo | 360 | 605 | -2.5 | N24E19 | 1852-1908 | M5.3 | 0 | ---- | 0 | ---- | |
| 20000122 | 00:23 | A | 226.10 | 2.20 | 1 | EJ | 17.4 | 17:50 | 400 | 20000118 | 17:54 | Halo | 360 | 739 | -7.1 | S19E11 | 1707-1744 | M3.9 | 1 | 1719 | 3 | 1731 | |

| | | | | | | | | | | | | | | | | | | | | | | |
|---|---|---|---|---|---|---|---|---|---|---|---|---|---|---|---|---|---|---|---|---|---|---|
| 20000211 | 02:33 | W,A,S | 506.70 | 1.91 | 1 | EJ | 14.4 | 17:00 | 425 | 20000208 | 09:30 | Halo | 360 | 1079 | -35.3 | N25E26 | 0842-0918 | M1.3 | 1 | 0857 | 3 | 0905 | 5 |
| 20000211 | 23:28 | W,A,S | 638.30 | 4.05 | 1 | MC | 17.6 | *17:06 | 543 | 20000209 | 19:54 | Halo | 360 | 910 | -1.1 | S17W40 | 1915-2059 | C7.4 | 0 | ---- | 2 | 0845 | 12 |
| 20000214 | 07:18 | W,A,S | 606.90 | 2.07 | 1 | EJ | 5.2 | 12:30 | 550 | 20000212 | 04:31 | Halo | 360 | 1107 | -8.3 | N26W23 | 0351-0431 | M1.7 | 1 | 0406 | 3 | 0355 | |
| 20000220 | 21:00 | W,A,S | 471.90 | 3.22 | 1 | MC | 12.8 | *09:48 | 380 | 20000217 | 21:30 | Halo | 360 | 728 | -22.9 | S29E07 | 2017-2107 | M1.3 | 1 | 2025 | 3 | 2042 | |
| 20000406 | 16:27 | W,A,S | 641.50 | 4.25 | 1 | EJ | 4.9 | 21:20 | 570 | 20000404 | 16:32 | Halo | 360 | 1188 | 12.8 | N16W66 | 1512-1605 | C9.7 | 1 | 1525 | 1 | 1545 | |
| 20000604 | 14:53 | W,A,S | 647.60 | 3.69 | 1 | EJ | 9.1 | *00:00 | 470 | 20000602 | 10:30 | 21 | >54 | 442 | 13.1 | N10E23 | 0900-1031 | C2.4 | 0 | ---- | 0 | ---- | 13 |
| 20000608 | 09:04 | W,A,S | 868.00 | 4.88 | 1 | EJ | 2.9 | 12:00 | 760 | 20000606 | 15:54 | Halo | 360 | 1119 | 1.5 | N20E18 | 1458-1540 | X2.3 | 1 | 1516 | 3 | 1520 | |
| 20000623 | 13:00 | W,A,S | 489.00 | 3.39 | 1 | MC | 19.3 | *08:18 | 500 | 20000620 | 09:10 | 204 | 65 | 636 | 6.1 | S30W30 | 0800-0838 | C2.1 | 0 | ---- | 0 | ---- | |
| 20000710 | 06:00 | A,S | 485.00 | 1.84 | 1 | EJ | 19.5 | *01:30 | 440 | 20000707 | 10:26 | Halo | 360 | 453 | 10.8 | N17E10 | 0842-1011 | C5.6 | 0 | ---- | 0 | ---- | |
| 20000711 | 11:22 | A | 518.40 | 2.05 | 1 | EJ | 11.4 | 22:48 | 520 | 20000708 | 23:50 | 354 | 161 | 483 | -7.2 | N18W12 | 2258-0124 | C4.0 | 0 | ---- | 1 | 2230 | |
| 20000713 | 09:49 | W,A,S | 632.40 | 2.03 | 1 | EJ | 7.2 | 17:00 | 600 | 20000710 | 21:50 | 67 | 289 | 1352 | 35.0 | N18E49 | 2105-2227 | M5.7 | 1 | 2121 | 1 | 2200 | |
| 20000714 | 15:39 | W,A,S | 833.00 | 2.40 | 3 | MC | 15.2 | *06:48 | 650 | 20000711 | 13:27 | Halo | 360 | 1078 | -42.9 | N18E27 | 1212-1335 | X1.0 | 0 | ---- | 1 | 1300 | |
| 20000715 | 14:18 | W,A,S | 1350.00 | 7.50 | 3 | MC | 6.8 | 21:06 | 990 | 20000714 | 10:54 | Halo | 360 | 1674 | -96.1 | N22W07 | 1003-1043 | X5.7 | 1 | 1017 | 3 | 1030 | |
| 20000719 | 14:50 | W,A,S | 639.10 | 2.97 | 1 | EJ | 17.1 | *07:56 | 570 | 20000717 | 08:54 | 95 | 116 | 788 | 30.7 | S10E36 | 0824-0848 | C5.3 | 0 | ---- | 0 | ---- | 14 |
| 20000726 | 03:02 | W,A,S | 382.90 | 2.50 | 1 | S | ---- | ----- | --- | 20000722 | 11:54 | 259 | 229 | 1230 | -12.4 | N14W56 | 1117-1202 | M3.7 | 1 | 1125 | 1 | 1145 | |
| 20000726 | 18:58 | W,A,S | 421.00 | 2.22 | 1 | EJ | 13.5 | *08:28 | 350 | 20000723 | 05:30 | 161 | 181 | 631 | -20.4 | S13W05 | 0411----- | EP | 0 | ---- | 0 | ---- | |
| 20000728 | 06:39 | W,A,S | 490.20 | 2.01 | 1 | MC | 15.3 | 21:06 | 471 | 20000725 | 03:30 | Halo | 360 | 528 | -5.8 | N06W08 | 0243-0254 | M8.0 | 1 | 0249 | 2 | 0930 | 3 |
| 20000728 | 09:09 | A,S | 452.50 | 1.68 | 1 | S | ---- | ----- | --- | 20000725 | 04:40 | ---- | --- | --- | ---- | S13W71 | 0440-0504 | M3.7 | - | ---- | - | ---- | 5,15 |
| 20000810 | 05:10 | W,A | 404.30 | 2.01 | 1 | EJ | 13.8 | 19:00 | 430 | 20000806 | 23:06 | 234 | 40 | 597 | -7.0 | S24W15 | 2236----- | --- | 0 | ---- | 0 | ---- | 6,16 |
| 20000811 | 18:51 | W,A,S | 562.80 | 1.41 | 1 | MC | 11.2 | *06:06 | 567 | 20000809 | 16:30 | Halo | 360 | 702 | 2.8 | N20E12 | 1519----- | EP | 0 | ---- | 0 | ---- | 17 |
| 20000814 | 21:36 | A,S | 381.89 | 2.60 | 2 | EJ | 4.4 | *02:00 | 560 | 20000812 | 14:54 | 283 | 161 | 876 | 10.6 | N13W46 | 1348-1707 | C3.2 | 0 | ---- | 0 | ---- | |
| 20000906 | 17:04 | W,A,S | 538.20 | 2.46 | 1 | EJ | 8.9 | *02:00 | 450 | 20000904 | 06:06 | 327 | 145 | 849 | 7.6 | N30W37 | 0448----- | EP | 0 | ---- | 2 | 0325 | 3,7 |
| 20000915 | 04:28 | W,A,S | 376.80 | 4.66 | 1 | EJ | 14.9 | 19:22 | 380 | 20000912 | 11:54 | Halo | 360 | 1550 | 58.2 | S19W06 | 1131-1313 | M1.0 | 1 | 1133 | 3 | 1200 | |
| 20000917 | 17:00 | W,A | 1022.70 | 4.51 | 1 | MC | 8.9 | *01:54 | 760 | 20000916 | 05:18 | Halo | 360 | 1215 | -12.3 | N14W07 | 0406-0448 | M5.9 | 1 | 0417 | 3 | 0430 | |
| 20001003 | 01:00 | W,A | 461.80 | 2.33 | 1 | MC | 16.1 | 17:06 | 409 | 20000929 | 21:50 | 114 | 274 | 173 | 8.8 | S22E53 | 2011----- | --- | 0 | ---- | 0 | ---- | 6,7 |
| 20001005 | 03:23 | W,A,S | 538.10 | 3.91 | 1 | EJ | 9.8 | 13:13 | 530 | 20001002 | 03:50 | Halo | 360 | 525 | -4.9 | S09E07 | 0248-0313 | C4.1 | 0 | ---- | 0 | ---- | |
| 20001012 | 22:36 | W,A,S | 525.80 | 2.83 | 1 | MC | 19.8 | *18:24 | 395 | 20001009 | 23:50 | Halo | 360 | 798 | -9.8 | N01W14 | 2319-0021 | C6.7 | 1 | 2338 | 0 | ---- | |
| 20001028 | 06:45 | W,A,S | 410.00 | 1.92 | 1 | S | ---- | ----- | --- | 20001024 | 08:26 | Halo | 360 | 800 | -0.6 | S23E70 | 0804-1144 | C2.3 | 0 | ---- | 0 | ---- | |
| 20001028 | 09:35 | W,A,S | 393.50 | 2.36 | 1 | MC | 13.7 | 23:18 | 375 | 20001025 | 08:26 | Halo | 360 | 770 | 17.4 | N09W63 | 0845-1521 | C4.0 | 0 | ---- | 3 | 0930 | |
| 20001106 | 09:20 | W,A,S | 617.00 | 3.10 | 1 | MC | 13.8 | 23:06 | 535 | 20001103 | 18:26 | Halo | 360 | 291 | 16.4 | N02W02 | 1835-2006 | C3.2 | 0 | ---- | 1 | 1835 | 18 |
| 20001110 | 06:20 | W,S | 924.80 | 4.40 | 1 | EJ | 11.7 | 18:00 | 850 | 20001108 | 23:06 | 271 | 170 | 1738 | 69.9 | N10W77 | 2242-0005 | M7.4 | 0 | ---- | 3 | 2320 | |
| 20001126 | 05:30 | W,A,S | 470.50 | 2.34 | 1 | EJ | 2.5 | 08:00 | 475 | 20001124 | 05:30 | Halo | 360 | 1289 | 2.1 | N20W05 | 0455-0508 | X2.0 | 1 | 0502 | 3 | 0510 | |
| 20001126 | 11:40 | W,A,S | 518.40 | 2.41 | 1 | EJ | 19.8 | *07:26 | 610 | 20001125 | 01:31 | Halo | 360 | 2519 | -5.0 | N07E50 | 0059-0201 | M8.2 | 1 | 0107 | 1 | 0125 | |
| 20001128 | 05:30 | W,A,S | 603.00 | 2.45 | 1 | EJ | 13.5 | 19:00 | 560 | 20001126 | 17:06 | Halo | 360 | 980 | 5.8 | N18W38 | 1634-1656 | X4.0 | 1 | 1655 | 1 | 1700 | |
| 20010113 | 02:31 | W,A,S | 386.70 | 1.34 | 1 | EJ | 6.5 | 09:00 | 420 | 20010110 | 00:54 | Halo | 360 | 832 | 7.9 | N13E36 | 0023-0140 | C5.9 | 1 | 0042 | 2 | 0330 | 9 |
| 20010117 | 16:24 | W,A,S | 406.10 | 3.70 | 1 | EJ | 8.2 | *00:37 | 350 | 20010114 | 06:30 | 327 | 134 | 945 | 24.7 | N48W80 | 0659-0703 | C1.4 | 0 | ---- | 2 | 2030 | 19 |
| 20010123 | 10:49 | W,A,S | 617.30 | 3.04 | 1 | EJ | 22.2 | *09:00 | 425 | 20010120 | 21:30 | Halo | 360 | 1507 | -41.1 | S07E46 | 2106-2132 | M7.7 | 1 | 2112 | 3 | 2130 | |
| 20010131 | 08:36 | W,A,S | 488.20 | 4.24 | 1 | S | ---- | ----- | --- | 20010128 | 15:54 | Halo | 360 | 916 | 3.5 | S04W59 | 1540-1624 | M1.5 | 0 | ---- | 3 | 1545 | 20 |
| 20010220 | 02:30 | W,A,S | 252.64 | 2.49 | 2 | EJ | 9.8 | 12:20 | 320 | 20010215 | 13:54 | Halo | 360 | 625 | 4.0 | N07E12 | 1308-1518 | B8.8 | 0 | ---- | 0 | ---- | |
| 20010303 | 11:30 | W,A,S | 516.90 | 3.27 | 1 | EJ | 17.7 | *05:11 | 460 | 20010228 | 14:50 | 252 | 232 | 313 | 1.9 | S17W05 | 1322-1340 | B4.2 | 0 | ---- | 0 | ---- | |
| 20010319 | 11:30 | W,A,S | 441.00 | 2.97 | 1 | MC | 11.8 | 23:18 | 420 | 20010316 | 03:50 | 7 | 281 | 271 | 2.6 | N11W09 | --------- | --- | 0 | ---- | 0 | ---- | 21 |
| 20010322 | 14:00 | W,A,S | 382.70 | 1.32 | 1 | EJ | 8.5 | 22:30 | 390 | 20010319 | 05:26 | Halo | 360 | 389 | -2.4 | S20W00 | 0412----- | EP | 0 | ---- | 0 | ---- | |
| 20010327 | 02:02 | W,A,S | 385.50 | 6.43 | 1 | S | ---- | ----- | --- | 20010324 | 20:50 | Halo | 360 | 906 | 42.9 | N15E22 | 1935-2047 | M1.7 | 0 | ---- | 0 | ---- | |
| 20010327 | 18:05 | W,A,S | 559.30 | 2.31 | 1 | EJ | 5.5 | 23:32 | 630 | 20010325 | 17:06 | Halo | 360 | 677 | -12.2 | N16E25 | 1625-1710 | C9.0 | 0 | ---- | 0 | ---- | |
| 20010331 | 01:14 | W,A,S | 565.80 | 12.02 | 1 | EJ | 3.8 | 05:00 | 650 | 20010329 | 10:26 | Halo | 360 | 942 | 3.5 | N20W19 | 0957-1032 | X1.7 | 1 | 1003 | 3 | 1012 | |
| 20010404 | 14:40 | W,A,S | 834.20 | 5.75 | 1 | MC | 6.2 | 20:54 | 740 | 20010402 | 22:06 | 261 | 244 | 2505 | 108.5 | N19W72 | 2132-2203 | X20.0 | 1 | 2152 | 3 | 2205 | |
| 20010407 | 17:58 | W,A,S | 607.00 | 3.48 | 1 | S | ---- | ----- | --- | 20010405 | 17:06 | Halo | 360 | 1390 | -21.8 | S24E50 | 1657-1814 | M5.1 | 0 | ---- | - | ---- | 22 |
| 20010408 | 11:20 | W,A,S | 698.50 | 5.81 | 1 | EJ | 7.7 | 19:00 | 750 | 20010406 | 19:30 | Halo | 360 | 1270 | -57.3 | S21E31 | 1910-1931 | X5.6 | 0 | ---- | 3 | 1935 | |
| 20010411 | 14:12 | W,A,S | 685.60 | 2.40 | 1 | EJ | 8.3 | 22:30 | 720 | 20010409 | 15:54 | Halo | 360 | 1192 | 1.3 | S21W04 | 1520-1600 | M7.9 | 1 | 1527 | 3 | 1553 | 5 |
| 20010411 | 16:19 | W,A,S | 809.80 | 2.65 | 1 | MC | 15.6 | *07:54 | 670 | 20010410 | 05:30 | Halo | 360 | 2411 | 211.6 | S23W09 | 0506-0542 | X2.3 | 1 | 0513 | 3 | 0524 | |
| 20010413 | 07:25 | W,A,S | 856.41 | 1.93 | 2 | EJ | 2.7 | 10:09 | 830 | 20010411 | 13:31 | Halo | 360 | 1103 | -13.0 | S22W27 | 1256-1349 | M2.3 | 1 | 1303 | 1 | 1315 | |
| 20010418 | 00:52 | W,A | 599.00 | 4.85 | 1 | EJ | 24.9 | *01:44 | 470 | 20010415 | 14:06 | 245 | 167 | 1199 | -35.9 | S20W85 | 1319-1355 | X14.4 | 1 | 1348 | 3 | 1405 | 23 |
| 20010421 | 15:30 | W,A,S | 384.20 | 2.09 | 1 | MC | 9.4 | *00:54 | 395 | 20010419 | 12:30 | 291 | 129 | 392 | 13.8 | N20W20 | 0812----- | EP | 0 | ---- | 0 | ---- | |
| 20010428 | 05:02 | W,A,S | 930.30 | 5.05 | 1 | MC | 20.9 | *01:54 | 640 | 20010426 | 12:30 | Halo | 360 | 1006 | 21.1 | N20W05 | 1126-1319 | M1.5 | 1 | 1335 | 3 | 1240 | |
| 20010527 | 14:45 | W,A,S | 615.60 | 2.72 | 1 | MC | 21.1 | *11:54 | 475 | 20010525 | 04:06 | 354 | 91 | 569 | 12.8 | N32W20 | 0312----- | EP | 0 | ---- | 0 | ---- | 24 |
| 20010618 | 02:10 | W,A,S | 337.50 | 4.91 | 1 | EJ | 22.6 | 23:10 | 390 | 20010615 | 10:31 | 185 | 119 | 1090 | 9.7 | S26E41 | 1001-1020 | M6.3 | 1 | 1006 | 1 | 1050 | 25 |
| 20010805 | 12:30 | W,A,S | 355.78 | 1.30 | 2 | EJ | 10.4 | 22:57 | 450 | 20010802 | 20:52 | 59 | 123 | 453 | 7.7 | N36E36 | 1855----- | EP | 0 | ---- | 0 | ---- | |
| 20010812 | 11:10 | W,A,S | 422.70 | 3.16 | 1 | EJ | 3.1 | 14:16 | 360 | 20010809 | 10:30 | 285 | 175 | 479 | 4.4 | N11W14 | 0800----- | DIM | 0 | ---- | 2 | 0100 | 3,26 |
| 20010817 | 11:00 | W,A,S | 519.20 | 3.15 | 1 | EJ | 11.6 | 22:38 | 540 | 20010814 | 16:01 | Halo | 360 | 618 | -4.8 | N16W36 | 1130-1242 | C2.3 | 0 | ---- | 0 | ---- | |
| 20010827 | 19:40 | W,A,S | 619.10 | 3.32 | 1 | S | ---- | ----- | --- | 20010825 | 16:50 | Halo | 360 | 1433 | -46.8 | S17E34 | 1623-1704 | X5.3 | 1 | 1632 | 3 | 1650 | |

| | | | | | | | | | | | | | | | | | | | | | | | | |
|---|---|---|---|---|---|---|---|---|---|---|---|---|---|---|---|---|---|---|---|---|---|---|---|---|
| 20010830 | 14:00 | W,S | 551.90 | 2.43 | 1 | EJ | 6.0 | 20:00 | 440 | 20010827 | 17:26 | 258 | 71 | 408 | -0.2 | S10W71 | 1648----- | --- | 0 | ---- | 0 | ---- | 27 |
| 20010913 | 02:30 | W | 449.10 | 1.36 | 1 | S | ---- | ----- | --- | 20010911 | 01:54 | 289 | 78 | 304 | -2.8 | NWlimb | 0113----- | DIM | 0 | ---- | 0 | ---- | 6,7 |
| 20010914 | 02:00 | W,A,S | 471.30 | 3.32 | 1 | EJ | 5.4 | 07:24 | 440 | 20010911 | 14:54 | Halo | 360 | 791 | 13.2 | N13E35 | 1400-1508 | C3.2 | 0 | ---- | 2 | 1000 | 3 |
| 20010925 | 20:18 | W,A | 835.10 | 6.57 | 1 | EJ | 7.2 | *03:30 | 600 | 20010924 | 10:30 | Halo | 360 | 2402 | 54.1 | S16E23 | 0932-1109 | X2.6 | 0 | ---- | 3 | 1045 | |
| 20010929 | 09:25 | W,A,S | 735.00 | 3.37 | 1 | EJ | 15.6 | *01:03 | 590 | 20010927 | 08:06 | 224 | 138 | 669 | 21.3 | S15W40 | 0855-1406 | M1.0 | 0 | ---- | 3 | 0815 | |
| 20010930 | 19:15 | W,A,S | 1021.20 | 13.17 | 1 | EJ | 2.4 | 21:40 | 520 | 20010928 | 08:54 | Halo | 360 | 846 | -6.9 | N10E18 | 0810-0910 | M3.3 | 1 | 0817 | 0 | ---- | |
| 20011011 | 16:50 | W,A,S | 581.50 | 3.07 | 1 | EJ | 11.4 | *04:16 | 590 | 20011009 | 11:30 | Halo | 360 | 973 | -41.5 | S28E08 | 1046-1149 | M1.4 | 1 | 1054 | 3 | 1115 | |
| 20011021 | 16:40 | W,A,S | 636.40 | 5.64 | 1 | EJ | 6.3 | 22:57 | 640 | 20011019 | 16:50 | Halo | 360 | 901 | -0.7 | N15W29 | 1613-1643 | X1.6 | 1 | 1624 | 3 | 1645 | |
| 20011025 | 09:00 | W,A,S | 358.30 | 3.87 | 1 | EJ | 12.0 | 21:00 | 410 | 20011022 | 15:06 | Halo | 360 | 1336 | -8.0 | S21E18 | 1427-1531 | M6.7 | 1 | 1453 | 1 | 1515 | 5 |
| 20011028 | 03:10 | W,A,S | 586.20 | 2.59 | 1 | EJ | 36.8 | *16:00 | 410 | 20011025 | 15:26 | Halo | 360 | 1092 | -1.4 | S16W21 | 1442-1528 | X1.3 | 1 | 1455 | 3 | 1530 | 23 |
| 20011031 | 13:47 | W,A,S | 400.50 | 2.59 | 1 | MC | 7.5 | 21:18 | 340 | 20011027 | 08:05 | ---- | --- | ---- | ---- | N06W02 | 0805-0858 | C6.3 | 0 | ---- | 0 | ---- | 2,28 |
| 20011106 | 01:45 | W,A,S | 742.00 | 2.10 | 3 | EJ | 4.2 | 06:00 | 700 | 20011104 | 16:35 | Halo | 360 | 1810 | -63.4 | N06W18 | 1603-1657 | X1.0 | 1 | 1610 | 3 | 1630 | |
| 20011119 | 18:15 | W,A,S | 628.30 | 2.84 | 1 | EJ | 8.8 | *03:00 | 490 | 20011117 | 05:30 | Halo | 360 | 1379 | -22.5 | S13E42 | 0449-0611 | M2.8 | 1 | 0450 | 2 | 1700 | |
| 20011124 | 06:00 | W,A,S | 1023.00 | 5.67 | 1 | MC | 9.8 | 15:48 | 730 | 20011122 | 23:30 | Halo | 360 | 1437 | -12.9 | S17W36 | 2232-0006 | M9.9 | 1 | 2231 | 3 | 2240 | |
| 20011229 | 05:15 | W,A,S | 527.60 | 2.20 | 1 | EJ | 20.2 | *01:30 | 390 | 20011226 | 05:30 | 281 | 212 | 1446 | -39.9 | N08W54 | 0432-0647 | M7.1 | 1 | 0502 | 3 | 0520 | |
| 20011230 | 20:00 | W,A,S | 670.40 | 2.33 | 1 | S | ---- | ----- | --- | 20011228 | 20:30 | Halo | 360 | 2216 | 6.9 | S24E90 | 2002-2132 | X3.4 | 0 | ---- | 3 | 2035 | |
| 20020110 | 16:30 | W,A,S | 634.28 | 2.10 | 2 | S | ---- | ----- | --- | 20020108 | 17:54 | Halo | 360 | 1794 | 81.4 | SElimb | 1814-2039 | C9.6 | 0 | ---- | 3 | 1830 | 25 |
| 20020117 | 05:50 | W,A,S | 277.20 | 1.62 | 1 | S | ---- | ----- | --- | 20020114 | 05:35 | Halo | 360 | 1492 | 52.3 | S28W90 | 0529-0825 | M4.4 | 1 | 0608 | 1 | 0625 | |
| 20020131 | 21:38 | W,A,S | 388.00 | 2.93 | 1 | EJ | 13.4 | *11:00 | 350 | 20020128 | 10:54 | 143 | 62 | 524 | 35.0 | S30E19 | 0936----- | EP | 0 | ---- | 0 | ---- | 29 |
| 20020217 | 03:30 | W,A,S | 324.70 | 2.80 | 1 | S | ---- | ----- | --- | 20020213 | 20:30 | 112 | 204 | 735 | -8.3 | SElimb | 1946----- | DIM | 0 | ---- | 2 | ---- | 3 |
| 20020228 | 05:10 | W,A,S | 382.90 | 1.40 | 1 | EJ | 11.7 | 16:49 | 410 | 20020224 | 15:30 | 260 | 54 | 231 | 5.2 | S18W44 | 1435-1500 | C4.4 | 0 | ---- | 0 | ---- | |
| 20020315 | 18:28 | A,S | 359.00 | 1.40 | 3 | S | ---- | ----- | --- | 20020312 | 23:54 | 112 | 82 | 535 | -5.7 | S22E90 | 2301-2324 | M1.5 | 1 | 2319 | 0 | ---- | |
| 20020318 | 13:13 | W,A,S | 577.00 | 9.27 | 1 | MC | 33.7 | *22:54 | 370 | 20020315 | 23:06 | Halo | 360 | 957 | -17.4 | S08W03 | 2209-0042 | M2.2 | 0 | ---- | 1 | 2245 | 30 |
| 20020320 | 13:22 | W,A,S | 814.20 | 1.31 | 1 | EJ | 27.1 | *16:28 | 440 | 20020318 | 02:54 | Halo | 360 | 989 | -2.9 | S10W30 | 0216-0400 | M1.0 | 0 | ---- | 1 | 0255 | 23 |
| 20020323 | 11:24 | W,A,S | 516.30 | 2.23 | 1 | MC | 16.4 | *03:48 | 440 | 20020320 | 17:54 | Halo | 360 | 603 | -15.8 | S17W20 | 1544-1919 | C3.3 | 0 | ---- | 0 | ---- | |
| 20020325 | 01:16 | W,A,S | 636.53 | 1.11 | 1 | EJ | 4.7 | 06:00 | 440 | 20020320 | 23:54 | 242 | 160 | 1075 | -0.2 | S19W60 | 2346-0046 | C5.7 | 0 | ---- | 0 | ---- | |
| 20020417 | 11:01 | W,A,S | 516.80 | 5.12 | 1 | MC | 17.3 | *04:18 | 480 | 20020415 | 03:50 | Halo | 360 | 720 | 2.1 | S15W01 | 0305-0506 | M1.2 | 0 | ---- | 1 | 0335 | |
| 20020419 | 08:02 | W,A,S | 748.00 | 1.72 | 1 | MC | 27.8 | *11:48 | 510 | 20020417 | 08:26 | Halo | 360 | 1240 | -19.8 | S14W34 | 0746-0957 | M2.6 | 1 | 0808 | 3 | 0830 | 1,30 |
| 20020423 | 05:12 | W,A,S | 639.00 | 3.24 | 1 | S | ---- | ----- | --- | 20020421 | 01:27 | Halo | 360 | 2393 | -1.4 | S14W84 | 0043-0238 | X1.5 | 1 | 0119 | 3 | 0130 | |
| 20020510 | 10:29 | A,S | 431.30 | 2.55 | 1 | EJ | 11.5 | 22:00 | 350 | 20020507 | 04:06 | Halo | 360 | 720 | 158.2 | S10E25 | 0337-0407 | M1.4 | 0 | ---- | 0 | ---- | |
| 20020511 | 10:30 | W,A,S | 481.70 | 3.31 | 1 | EJ | 2.5 | 13:00 | 430 | 20020508 | 13:50 | Halo | 360 | 614 | 78.9 | S12W07 | 1258-1359 | C4.2 | 0 | ---- | 0 | ---- | |
| 20020518 | 19:51 | W,A,S | 546.10 | 5.29 | 1 | MC | 8.1 | *03:54 | 458 | 20020516 | 00:50 | Halo | 360 | 600 | -6.6 | S23E15 | 0011-0118 | C4.5 | 1 | 0028 | 1 | 0150 | |
| 20020520 | 03:40 | W,A,S | 541.90 | 1.67 | 1 | EJ | 11.3 | 15:00 | 450 | 20020517 | 01:27 | 145 | 45 | 461 | 5.5 | S20E14 | 0023----- | EP | 0 | ---- | 0 | ---- | 6 |
| 20020521 | 21:14 | W,A | 326.30 | 1.93 | 1 | EJ | 2.8 | *00:00 | 390 | 20020518 | 11:50 | 154 | 46 | 614 | 45.6 | S17E36 | 1042-1209 | C3.0 | 0 | ---- | 0 | ---- | 16 |
| 20020523 | 10:40 | W,A,S | 735.80 | 3.46 | 1 | MC | 12.7 | 23:24 | 730 | 20020522 | 03:50 | Halo | 360 | 1557 | -10.4 | S30W34 | 0318-0502 | C5.0 | 0 | ---- | 2 | 0410 | |
| 20020530 | 02:15 | W,A,S | 689.40 | 8.28 | 1 | EJ | 4.9 | 07:09 | 500 | 20020527 | 13:27 | 49 | 161 | 1106 | 3.8 | N22E15 | 1236-1351 | C3.7 | 0 | ---- | 0 | ---- | 20 |
| 20020717 | 15:50 | W,A,S | 493.00 | 4.15 | 1 | EJ | 20.2 | *12:00 | 450 | 20020715 | 21:30 | 14 | 188 | 1300 | -7.3 | N19W01 | 2103-2148 | M1.8 | 0 | ---- | 3 | 2115 | |
| 20020719 | 14:40 | A,S | 556.40 | 2.87 | 1 | EJ | 18.3 | *09:00 | 700 | 20020718 | 08:06 | Halo | 360 | 1099 | -30.2 | N19W30 | 0724-0749 | X1.8 | 1 | 0744 | 1 | 0755 | |
| 20020725 | 12:59 | A,S | 584.10 | 2.47 | 1 | S | ---- | ----- | --- | 20020723 | 00:42 | Halo | 360 | 2285 | --- | S13E72 | 0018-0047 | X4.8 | 1 | 0029 | 3 | 0050 | |
| 20020729 | 12:40 | A,S | 527.10 | 4.32 | 1 | EJ | 19.3 | *08:00 | 420 | 20020726 | 22:06 | Halo | 360 | 818 | -0.1 | S19E26 | 2051-2129 | M8.7 | 0 | ---- | 1 | 2227 | |
| 20020801 | 05:10 | W,A,S | 494.70 | 1.41 | 1 | MC | 6.7 | 11:54 | 454 | 20020729 | 12:07 | 13 | 236 | 562 | -4.3 | S10W10 | 1027-1113 | M4.7 | 0 | ---- | 1 | 1210 | 31 |
| 20020801 | 23:05 | W,A,S | 494.70 | 1.41 | 1 | MC | 8.3 | *07:24 | 460 | 20020729 | 23:30 | 351 | 130 | 360 | 2.8 | N12W16 | 1823----- | EP | 0 | ---- | 0 | ---- | 32 |
| 20020818 | 18:40 | W,A,S | 671.30 | 7.75 | 1 | EJ | 10.1 | *04:45 | 550 | 20020816 | 12:30 | Halo | 360 | 1585 | -67.1 | S14E20 | 1132-1307 | M5.2 | 1 | 1144 | 3 | 1220 | |
| 20020820 | 13:50 | W,A | 535.20 | 1.21 | 1 | EJ | 10.2 | *00:00 | 425 | 20020818 | 21:54 | 203 | 140 | 682 | 1.9 | S12W19 | 2112-2137 | M2.2 | 1 | 2136 | 0 | ---- | |
| 20020826 | 11:10 | W,A,S | 372.04 | 2.45 | 2 | S | ---- | ----- | --- | 20020824 | 01:27 | Halo | 360 | 1913 | 43.7 | S02W81 | 0049-0131 | X3.1 | 1 | 0109 | 3 | 0145 | |
| 20020907 | 16:20 | W,A,S | 897.70 | 6.27 | 1 | EJ | 19.7 | *12:00 | 460 | 20020905 | 16:54 | Halo | 360 | 1748 | 43.0 | N09E28 | 1618-1735 | C5.2 | 1 | 1635 | 3 | 1655 | |
| 20020930 | 07:50 | W,A,S | 329.60 | 1.32 | 1 | MC | 14.8 | 22:36 | 381 | 20020928 | 11:06 | 70 | 79 | 678 | -6.2 | N12E33 | 1035-1121 | C3.4 | 0 | ---- | 0 | ---- | 6 |
| 20021002 | 22:08 | W,A,S | 511.80 | 3.89 | 1 | EJ | 4.5 | *02:37 | 480 | 20020930 | 01:31 | 202 | 124 | 307 | -1.7 | S12W16 | 2336----- | EP | 0 | ---- | 0 | ---- | 6 |
| 20021109 | 18:27 | W,A,S | 422.20 | 1.89 | 1 | EJ | 13.1 | *07:30 | 370 | 20021106 | 06:06 | 162 | 102 | 485 | -6.3 | S13E13 | 0505-0614 | C7.2 | 1 | 0538 | 0 | ---- | |
| 20021126 | 21:47 | W,A,S | 610.90 | 2.98 | 1 | EJ | 20.9 | *18:38 | 530 | 20021124 | 20:30 | Halo | 360 | 1077 | 20.5 | N20E35 | 1912----- | EP | 1 | 2006 | 1 | 2005 | |
| 20021224 | 13:20 | A,S | 531.20 | 1.59 | 1 | S | ---- | ----- | --- | 20021222 | 03:30 | 328 | 272 | 1071 | -10.1 | N23W42 | 0214-0312 | M1.1 | 1 | 0248 | 1 | 0420 | |
| 20030214 | 17:39 | A,S | 576.00 | 1.00 | 3 | S | ---- | ----- | --- | 20030212 | 02:30 | 267 | 118 | 355 | -1.0 | S05W43 | 0142-0202 | C8.7 | 1 | 0151 | 0 | ---- | |
| 20030320 | 04:20 | W,A,S | 528.60 | 1.22 | 2 | MC | 7.6 | 11:54 | 650 | 20030318 | 12:30 | 263 | 209 | 1601 | -13.3 | S15W46 | 1151-1220 | X1.5 | 1 | 1216 | 3 | 1225 | |
| 20030408 | 00:19 | W,A,S | 368.29 | 3.72 | 2 | S | ---- | ----- | --- | 20030404 | 21:19 | 291 | 89 | 487 | -2.4 | S11W40 | 1905-2040 | M1.9 | 1 | 2132 | 0 | ---- | |
| 20030424 | 18:19 | W,S | 625.00 | 1.20 | 3 | S | ---- | ----- | --- | 20030421 | 13:36 | 296 | 163 | 784 | -4.2 | N18E02 | 1254-1314 | M2.8 | 1 | 1305 | 0 | ---- | 20,33 |
| 20030529 | 11:52 | W,A,S | 677.86 | 1.57 | 2 | EJ | 1.1 | 13:00 | 650 | 20030527 | 23:50 | Halo | 360 | 964 | -9.6 | S07W17 | 2256-2313 | X1.3 | 1 | 2306 | 1 | 2312 | |
| 20030529 | 18:36 | W,A,S | 905.90 | 2.17 | 1 | EJ | 7.4 | *02:00 | 600 | 20030528 | 00:50 | Halo | 360 | 1366 | 25.9 | S07W20 | 0017-0039 | X3.6 | 1 | 0026 | 1 | 0100 | |
| 20030530 | 16:02 | W,A,S | 699.87 | 1.45 | 2 | EJ | 6.0 | 22:00 | 680 | 20030529 | 01:27 | Halo | 360 | 1237 | -22.3 | S06W37 | 0051-0112 | X1.2 | 1 | 0101 | 3 | 0110 | 33 |
| 20030618 | 04:44 | W,A,S | 576.10 | 2.21 | 1 | S | ---- | ----- | --- | 20030615 | 23:54 | Halo | 360 | 2053 | -0.9 | S07E80 | 2325-0025 | X1.3 | 1 | 2346 | 3 | 0000 | 34 |
| 20030620 | 07:56 | A,S | 556.49 | 1.42 | 2 | S | ---- | ----- | --- | 20030617 | 23:18 | Halo | 360 | 1813 | -2.9 | S07E55 | 2227-2312 | M6.8 | 1 | 2248 | 3 | 2250 | 33 |

| | | | | | | | | | | | | | | | | | | | | | |
|---|---|---|---|---|---|---|---|---|---|---|---|---|---|---|---|---|---|---|---|---|---|
| 20030817 | 13:40 | A,S | 528.48 | 2.29 | 2 | MC | 21.9 | *11:36 | 498 | 20030814 | 20:06 | Halo | 360 | 378 | 4.4 | S10E02 | 1712----- | Wave | 0 | ---- | 0 | ---- | |
| 20031024 | 14:47 | A,S | 655.82 | 2.85 | 2 | EJ | 7.7 | 22:28 | 580 | 20031021 | 03:54 | Halo | 360 | 1484 | -124.3 | Elimb | 0348----- | --- | 1 | 0347 | 1 | 0410 | |
| 20031026 | 07:49 | A,S | 625.58 | 1.19 | 3 | S | ---- | ----- | --- | 20031023 | 08:54 | 53 | 236 | 1406 | -16.5 | S21E88 | 0819-0849 | X5.4 | 1 | 0826 | 0 | ---- | |
| 20031026 | 18:35 | A,S | 607.02 | 1.11 | 2 | S | ---- | ----- | --- | 20031023 | 20:06 | 103 | 95 | 1136 | -26.1 | S17E84 | 1950-2014 | X1.1 | 0 | ---- | 0 | ---- | |
| 20031028 | 01:30 | A | 900.00 | 2.20 | 3 | EJ | 1.0 | 02:30 | 610 | 20031026 | 17:54 | 270 | 171 | 1537 | 4.8 | N02W38 | 1721-1921 | X1.2 | 1 | 1735 | 3 | 1745 | |
| 20031029 | 06:00 | A | 1825.00 | 13.65 | 3 | MC | 2.0 | 08:00 | 1500 | 20031028 | 11:30 | Halo | 360 | 2459 | -105.2 | S16E08 | 1100-1124 | X17.2 | 1 | 1102 | 3 | 1110 | |
| 20031030 | 16:20 | A | 1700.00 | 7.49 | 3 | MC | 9.7 | *02:00 | 1350 | 20031029 | 20:54 | Halo | 360 | 2029 | -146.5 | S15W02 | 2037-2101 | X10.0 | 1 | 2042 | 1 | 2055 | |
| 20031104 | 05:53 | W,A,S | 759.00 | 3.73 | 1 | S | ---- | ----- | --- | 20031102 | 17:30 | Halo | 360 | 2598 | -32.4 | S14W56 | 1703-1739 | X8.3 | 1 | 1714 | 3 | 1730 | |
| 20031106 | 18:56 | A,S | 335.01 | 2.20 | 2 | S | ---- | ----- | --- | 20031104 | 19:54 | Halo | 360 | 2657 | 434.8 | S19W83 | 1929-2006 | X28.0 | 1 | 1942 | 3 | 2000 | |
| 20031115 | 05:27 | A,S | 770.65 | 2.64 | 2 | S | ---- | ----- | --- | 20031113 | 09:30 | 49 | 217 | 1141 | 4.3 | N05E78 | 0903-1002 | M1.4 | 1 | 0922 | 1 | 0935 | |
| 20031120 | 07:28 | W,A,S | 666.40 | 3.95 | 1 | MC | 4.5 | 12:00 | 700 | 20031118 | 08:50 | Halo | 360 | 1660 | -3.3 | N00E18 | 0812-0859 | M3.9 | 1 | 0746 | 1 | 0746 | |
| 20031122 | 09:59 | A,S | 563.28 | 1.39 | 2 | S | ---- | ----- | --- | 20031120 | 08:06 | Halo | 360 | 669 | -23.8 | N01W08 | 0735-0753 | M9.6 | 0 | ---- | 0 | ---- | |
| 20040122 | 01:10 | A,S | 676.84 | 3.84 | 2 | EJ | 6.8 | 08:00 | 560 | 20040120 | 00:06 | Halo | 360 | 965 | 17.2 | S13W09 | 2346-0229 | C5.5 | 0 | ---- | 0 | ---- | |
| 20040123 | 14:20 | A | 531.86 | 1.34 | 2 | EJ | 8.7 | 23:00 | 490 | 20040121 | 04:54 | Halo | 360 | 762 | -13.7 | S19E29 | 0402-0530 | C1.2 | 0 | ---- | 0 | ---- | |
| 20040409 | 01:47 | A,S | 550.32 | 2.20 | 2 | S | ---- | ----- | --- | 20040406 | 13:31 | Halo | 360 | 1368 | 45.6 | S18E15 | 1230-1346 | M2.4 | 0 | ---- | 3 | 1305 | |
| 20040410 | 19:25 | A,S | 366.96 | 2.63 | 2 | S | ---- | ----- | --- | 20040408 | 10:30 | Halo | 360 | 1068 | -36.5 | S15W11 | 0953-1047 | C7.4 | 1 | 1006 | 3 | 1025 | |
| 20040412 | 17:35 | A,S | 399.98 | 2.73 | 2 | S | ---- | ----- | --- | 20040411 | 04:30 | 203 | 314 | 1645 | -77.6 | S16W46 | 0354-0435 | C9.6 | 0 | ---- | 3 | 0420 | |
| 20040722 | 09:45 | W,A,S | 458.70 | 2.30 | 2 | MC | 5.7 | 15:24 | 500 | 20040720 | 13:31 | Halo | 360 | 710 | 16.6 | N10E35 | 1222-1245 | M8.6 | 1 | 1235 | 0 | ---- | |
| 20040724 | 05:32 | W,A | 486.38 | 3.36 | 2 | MC | 7.3 | 12:48 | 600 | 20040722 | 08:30 | 193 | 132 | 899 | -12.6 | N04E10 | 0741-0808 | C5.3 | 0 | ---- | 0 | ---- | |
| 20040726 | 22:28 | W,A,S | 1100.50 | 9.47 | 2 | EJ | 5.5 | *04:00 | 870 | 20040725 | 14:54 | Halo | 360 | 1333 | 7.0 | N08W33 | 1419-1643 | M1.1 | 0 | ---- | 2 | 1500 | |
| 20040730 | 20:41 | A,S | 549.85 | 2.52 | 2 | S | ---- | ----- | --- | 20040729 | 12:06 | Halo | 360 | 1180 | 38.0 | N00W90 | 1142-1402 | C2.1 | 0 | ---- | 2 | 1320 | 33 |
| 20040801 | 02:02 | A,S | 500.65 | 2.04 | 2 | EJ | 13.5 | 15:33 | 460 | 20040731 | 05:54 | 259 | 197 | 1192 | 46.4 | N05W90 | 0516-0914 | C8.4 | 0 | ---- | 2 | 0710 | |
| 20040829 | 09:10 | W,S | 473.90 | 1.90 | 2 | MC | 9.5 | 18:42 | 393 | 20040826 | 12:30 | 206 | 119 | 184 | 5.5 | S09W34 | 1148----- | --- | 0 | ---- | 0 | ---- | |
| 20040913 | 19:29 | A,S | 505.00 | 1.10 | 3 | EJ | 23.4 | *18:51 | 550 | 20040912 | 00:36 | Halo | 360 | 1328 | 22.5 | N03E49 | 0004-0133 | M4.8 | 1 | 0023 | 3 | 0045 | |
| 20041107 | 02:22 | W,A,S | 303.68 | 2.56 | 2 | EJ | 2.3 | 04:40 | 360 | 20041104 | 09:54 | Halo | 360 | 653 | 6.3 | N09E28 | 0845-1004 | C6.3 | 0 | ---- | 0 | ---- | |
| 20041107 | 17:59 | W,A,S | 742.70 | 3.12 | 2 | MC | 9.4 | *03:24 | 674 | 20041104 | 23:30 | 338 | 293 | 1055 | -1.9 | N08E18 | 2253-2326 | M5.4 | 1 | 2222 | 0 | ---- | |
| 20041109 | 09:05 | W,A,S | 662.10 | 2.79 | 2 | MC | 11.8 | 20:54 | 800 | 20041106 | 02:06 | 351 | 214 | 1111 | 18.8 | N09E05 | 0140-0208 | M3.6 | 1 | 0044 | 3 | 0150 | |
| 20041109 | 18:24 | W,A,S | 785.61 | 4.01 | 2 | MC | 9.2 | *03:36 | 697 | 20041107 | 16:54 | Halo | 360 | 1759 | -19.7 | N09W17 | 1542-1615 | X2.0 | 1 | 1559 | 3 | 1625 | |
| 20041205 | 07:04 | A,S | 492.51 | 3.92 | 2 | S | ---- | ----- | --- | 20041203 | 00:26 | Halo | 360 | 1216 | -19.8 | N08W02 | 2344-0035 | M1.5 | 1 | 2353 | 3 | 0007 | 35 |
| 20041211 | 13:03 | A,S | 367.18 | 3.59 | 2 | EJ | 22.9 | *12:00 | 400 | 20041208 | 20:26 | Halo | 360 | 611 | -87.2 | N05W03 | 1934-2044 | C2.5 | 1 | 1945 | 1 | 2005 | |
| 20050116 | 09:27 | W | 586.40 | 1.37 | 2 | EJ | 4.5 | 14:00 | 520 | 20050115 | 06:30 | Halo | 360 | 2049 | -30.7 | N16E04 | 0554-0717 | M8.6 | 1 | 0555 | 3 | 0615 | |
| 20050117 | 07:15 | A | 465.33 | 4.60 | 2 | EJ | 6.2 | 13:30 | 780 | 20050115 | 23:06 | Halo | 360 | 2861 | -127.4 | N15W05 | 2225-2331 | X2.6 | 1 | 2234 | 3 | 2300 | |
| 20050118 | 18:23 | W | 741.04 | 1.32 | 2 | EJ | 6.3 | *00:39 | 940 | 20050117 | 09:54 | Halo | 360 | 2547 | -159.1 | N15W25 | 0942-0952 | X3.8 | 1 | 0916 | 3 | 0925 | 33 |
| 20050121 | 16:48 | A,S | 1067.50 | 7.50 | 3 | EJ | 7.2 | *00:00 | 850 | 20050120 | 06:54 | Halo | 360 | 3242 | 16.0 | N14W61 | 0636-0726 | X7.1 | 1 | 0644 | 3 | 0715 | |
| 20050217 | 21:59 | A,S | 443.69 | 1.19 | 2 | EJ | 17.0 | *15:00 | 540 | 20050213 | 11:06 | 151 | 151 | 584 | -13.0 | S11E09 | 1028-1051 | C2.7 | 1 | 1041 | 2 | 0130 | 3,33 |
| 20050515 | 02:19 | A,S | 840.77 | 6.91 | 2 | MC | 3.4 | 05:42 | 843 | 20050513 | 17:12 | Halo | 360 | 1689 | --- | N12E11 | 1613-1728 | M8.0 | 1 | 1644 | 3 | 1700 | |
| 20050520 | 03:34 | W,A | 394.35 | 1.89 | 2 | MC | 3.7 | 07:18 | 457 | 20050517 | 03:26 | 65 | 273 | 449 | 18.1 | S15W00 | 0231-0252 | M1.8 | 1 | 0236 | 1 | 0320 | 36 |
| 20050528 | 03:48 | A,S | 346.95 | 2.92 | 2 | EJ | 6.2 | 10:00 | 380 | 20050526 | 15:06 | Halo | 360 | 586 | -1.6 | S11E19 | 1310-1508 | B7.5 | 0 | ---- | 0 | ---- | |
| 20050529 | 09:15 | A,S | 526.88 | 2.46 | 2 | EJ | 1.0 | 10:15 | 490 | 20050526 | 21:26 | 144 | 199 | 420 | -1.8 | S08E11 | 2057-2204 | C8.6 | 0 | ---- | 0 | ---- | 33 |
| 20050612 | 06:59 | A,S | 296.60 | 9.88 | 2 | MC | 8.6 | 15:36 | 480 | 20050609 | 14:36 | 260 | 125 | 377 | -3.7 | N09E16 | 1328-1423 | C1.5 | 0 | ---- | 0 | ---- | 33,37 |
| 20050710 | 02:56 | A,S | 459.87 | 1.26 | 2 | EJ | 8.2 | 11:06 | 450 | 20050707 | 17:06 | Halo | 360 | 683 | -8.7 | N09E03 | 1607-1640 | M4.9 | 0 | ---- | 1 | 1630 | |
| 20050717 | 00:52 | W,A,S | 368.83 | 2.61 | 2 | MC | 14.4 | 15:18 | 428 | 20050715 | 15:00 | ---- | --- | --- | ---- | S15W29 | 1500----- | DSF | 0 | ---- | 0 | ---- | 38 |
| 20050801 | 06:09 | W,A,S | 541.22 | 1.92 | 2 | S | ---- | ----- | --- | 20050730 | 06:50 | Halo | 360 | 1968 | -102.6 | N12E60 | 0617-0701 | X1.3 | 1 | 0626 | 3 | 0740 | |
| 20050824 | 05:34 | W,A,S | 589.10 | 2.25 | 2 | S | ---- | ----- | --- | 20050822 | 17:30 | Halo | 360 | 2378 | 108.0 | S13W65 | 1646-1802 | M5.6 | 0 | ---- | 3 | 1715 | |
| 20050902 | 13:32 | W,A,S | 542.29 | 2.39 | 2 | EJ | 5.5 | 19:03 | 650 | 20050831 | 11:30 | Halo | 360 | 825 | 42.9 | N13W13 | 1026-1251 | C2.0 | 0 | ---- | 3 | 1140 | |
| 20050911 | 00:49 | W,A,S | 1350.00 | 7.30 | 3 | EJ | 4.2 | 05:00 | 1100 | 20050909 | 19:48 | Halo | 360 | 2257 | -128.6 | S12E67 | 1913-2036 | X6.2 | 1 | 1934 | 3 | 1945 | |
| 20050912 | 06:30 | W,A | 1205.78 | 2.48 | 3 | EJ | 13.5 | 20:00 | 750 | 20050910 | 21:52 | Halo | 360 | 1893 | -171.7 | S13E47 | 2130-2243 | X2.1 | 1 | 2137 | 3 | 2145 | |
| 20050915 | 08:25 | W,A,S | 715.76 | 5.19 | 3 | EJ | 6.0 | 14:24 | 790 | 20050913 | 20:00 | Halo | 360 | 1866 | 11.5 | S09E10 | 1942-2057 | X1.5 | 0 | ---- | 3 | 2020 | |
| 20060101 | 13:35 | W,A,S | 593.86 | 1.96 | 2 | EJ | 10.4 | *00:00 | 450 | 20051229 | 20:58 | 163 | 110 | 466 | 14.1 | S21E21 | 1950----- | DIM | 0 | ---- | 0 | ---- | |
| 20060413 | 11:42 | W,A,S | 328.36 | 1.87 | 2 | MC | 3.1 | 14:48 | 550 | 20060410 | 21:00 | ---- | --- | --- | ---- | S40E28 | --------- | --- | 0 | ---- | 0 | ---- | 33,39 |
| 20060709 | 20:39 | W,A,S | 403.44 | 2.33 | 2 | EJ | 10.7 | *07:20 | 390 | 20060706 | 08:54 | Halo | 360 | 911 | 8.2 | S09W34 | 0813-0851 | M2.5 | 1 | 0824 | 1 | 0845 | |
| 20060815 | 16:26 | W,A,S | 280.28 | 1.61 | 2 | S | ---- | ----- | --- | 20060811 | 19:31 | ---- | --- | --- | ---- | S13E37 | --------- | ---- | 0 | ---- | 0 | ---- | 40 |
| 20060818 | 15:42 | W,A,S | 493.52 | 2.24 | 2 | EJ | 7.3 | 23:00 | 360 | 20060816 | 07:31 | Halo | 360 | 563 | 5.1 | S01W19 | 0720-0819 | B3.4 | 0 | ---- | 0 | ---- | 41 |
| 20060819 | 10:51 | W,A,S | 407.95 | 1.33 | 2 | EJ | 13.2 | *00:00 | 450 | 20060816 | 16:30 | Halo | 360 | 888 | 1.9 | S16W08 | 1437-1922 | C3.6 | 0 | ---- | 3 | 1545 | |
| 20061208 | 04:01 | W,A,S | 831.54 | 4.33 | 2 | EJ | 7.0 | 11:00 | 630 | 20061206 | 20:12 | Halo | 360 | --- | --- | S05E64 | 1829-1900 | X6.5 | 1 | 1842 | 3 | 1900 | |
| 20061214 | 13:57 | W,A,S | 1016.19 | 8.53 | 2 | MC | 8.8 | 22:42 | 900 | 20061213 | 02:54 | Halo | 360 | 1774 | -61.4 | S06W23 | 0214-0257 | X3.4 | 1 | 0226 | 3 | 0245 | |
| 20061216 | 17:43 | W,A,S | 405.49 | 1.99 | 2 | EJ | 10.8 | *04:30 | 630 | 20061214 | 22:30 | Halo | 360 | 1042 | -0.4 | S06W46 | 2107-2226 | X1.5 | 1 | 2209 | 1 | 2230 | |

Notes:

Columns 1-6: Interplanetary (IP) shock arrival date, time (hh:mm), observing spacecraft (W-Wind, A-ACE, S-SOHO), shock speed (km/s), Alfvenic Mach numbers (Ma), and source of Mach number computation, C (1 - J. C. Kasper, 2 - A.Vinas, 3 – Approximation – see text)
Columns 7-10: ICME type (S = Sheath only, EJ = Ejecta, MC = magnetic cloud), shock standoff distance, ICME time (hh:mm, * next day), and speed (km/s).
Columns 11-16: White-light CME date, time (hh:mm), central position angle (CPA), width (degrees or Halo if 360 degrees), speed (km/s), acceleration (a in $ms^{-2}$). DG indicates CME data gap.
Columns 17-19: Solar flare location, duration (hhmm - hhmm), size in soft X-rays. EP = eruptive prominence, DIM = dimming, and Wave = EIT wave.
Columns 20-21: Metric type II code (C = 0 no metric emission, C =1 metric emission) and onset time (hhmm) of type II burst.
Columns 22-23: Decameter –hectometric (DH) type II code (C = 0 no radio emission, 1 = RAD2 (1-14 MHz), 2 = RAD1 (30 kHz - 1 MHz), 3 = RAD1+RAD2) and onset time (hhmm).
Column 24: Comment number (see below).

1: There are two shocks ahead of the MC. The shock-MC association is ambiguous.
2: Possible radio emission but may not be associated with this shock.
3: Wind/WAVES type II date is one day after the CME date.
4: CME solar source identification has less confidence, which is possibly associated with a large interconnecting loop between NOAA 8126 (N20E48) and NOAA 8124 (S20E10) on 1997/12/26 observed by SXT.
5: Ejecta identification has less confidence.
6: CME association has less confidence, i.e., other possible candidate CMEs exit.
7: Possible radio emission; association with this shock has less confidence.
8: Possible radio emission.
9: Wind/WAVES type II date is two days after the CME date.
10: There may be a faint halo not measured. The solar source is a C2.9 flare from AR8477 (S24W03) on 1999/03/07 at 03:54 UT.
11: MC event. Low confidence on CME association. The solar source may be correct, but the CME is too faint to be measured.
12: Wind/WAVES type II burst is on 2000/02/11.
13: CME has uncertain width (possible halo CME).
14: Radio emission may be associated with an earlier CME.
15: CME is unmeasured. Solar source is possibly associated with a M3.7 flare at AR9087 S13W71 on 2000/07/25 04:40 UT.
16: CME width is uncertain due to the compound CME.
17: Multiple eruptions from N20E12 and N11W11.
18: Multiple C3.2 flares are from N02W02 and N25E17.
19: CME is associated with a filament eruption.
20: Possible short ejecta (1-2 hours) after 24 hours of the shock, which may not be associated with this event.
21: CME may be backsided. Possible alternative candidate is the CME on 2001/03/15 at 22:26 UT.
22: Data gap from ~17:00 to ~ 19:00 UT, possible radio emission.
23: Ejecta association has less confidence. Ejecta after > 24 hours from the shock.
24: Alternative candidate is the CME on 2001/05/24 at 20:26 UT associated with the M1.2 flare on 05/24 at 19:30 UT from N07E29 (AR 9468).
25: Listed as a magnetic cloud-like event in http://lepmfi.gsfc.nasa.gov/mfi/MCL1.html
26: Clear radio emission but may not be associated with this shock.
27: Ejecta identification has less confidence. Ejecta is a marginal event.
28: MC event. CME is too faint to be measured. Solar source is a C6.3 flare from AR9678 (N06W02) on 2001/10/27 at 08:05 UT.
29: Possible complex event. Several low-beta intervals.
30: Possible complex event. MC onset after >24 hours from the shock.
31: Alternative candidate is the faint partial halo CME on 2002/07/29 at 00:30 UT from S22W11 (AR 10044).
32: It is possible that the previous CME candidate on 2002/07/29 at 12:07 UT is associated with the 2002/08/01 shock at 23:05 UT.
33: Shock identification has less confidence.
34: Proton temperature drop over a short interval (1 - 2 hrs) followed by high speed stream.

35: Metric type II burst is on 2004/12/02.
36: Alternative candidate is the CME on 2005/05/16 at 13:50 UT from N13W29.
37: Alternative candidate is the CME on 2005/06/07 at 10:24 UT from N25E20.
38: MC event. CME is too faint to be measured. Solar source is an EUV eruption on 2005/07/15 at 15:00 UT, location S15W29.
39: CME is too faint to measure. Solar source is possibly associated with a large faint arcade formation at S40E28 on 2006/04/10 ~21h by SXI. LASCO/C2 observed a faint spray-like front sweeping out the streamer at early 04/11. An alternate candidate is the CME on 04/10 at 06:06 UT whose south-east part may be backsided and south-west part may come from AR 10869 (S12W22).
40: CME is too faint to measure. Solar source is possibly associated with a cusp structure at S13E37 on 2006/08/11 ~19:31h by SXI. LASCO/C2 observed a large scale dimming overlaping the cusp arcade later.
41: Shock and CME source (flare association) identifications have less confidence.

Table 2. The annual number of IP shocks

| Year | # RQ | # RL | Total | # Shocks/day |
|---|---|---|---|---|
| 1996 | 3 | 0 | 3 | 0.01 |
| 1997 | 8 | 5 | 13 | 0.04 |
| 1998 | 11 | 9 | 20 | 0.07 |
| 1999 | 11 | 15 | 16 | 0.04 |
| 2000 | 12 | 23 | 35 | 0.10 |
| 2001 | 13 | 29 | 42 | 0.12 |
| 2002 | 12 | 23 | 35 | 0.10 |
| 2003 | 3 | 18 | 21 | 0.06 |
| 2004 | 5 | 13 | 18 | 0.05 |
| 2005 | 4 | 14 | 18 | 0.05 |
| 2006 | 4 | 5 | 9 | 0.03 |

Table 3. CME and shock speeds for all, RQ, and RL shocks according to the IP driver

| Shock | CME Speed (km/s) for | | | Shock Speed (km/s) for | | |
|---|---|---|---|---|---|---|
| Type (#) | MC (#) | EJ (#) | DL (#) | MC | EJ | DL |
| All (222) | 891 (57) | 941 (123) | 1308 (42) | 625 | 549 | 483 |
| RQ (76) | 436 (19) | 526 (46) | 743 (11) | 452 | 469 | 401 |
| RL (145) | 1118 (38) | 1189 (77) | 1512 (30) | 711 | 596 | 509 |

Table 4. Summary of properties of all, RQ and RL shocks

| Property | RQ | RL | All |
|---|---|---|---|
| Number of events | 76 | 145 | 222[a] |
| Average CME speed (km/s) | 535 | 1237 | 999 |
| Fraction of full halos (W = 360º)(%) | 40 | 72 | 61 |
| Fraction of wide CMEs (W ≥ 120º) (%) | 71 | 97 | 88 |
| Average width of non-halos (W < 120º) | 78 | 102 | 83 |
| Average acceleration ($m/s^2$) | 6.8 | -3.6 | -0.1 |
| Median flare size | C3.4 | M4.7 | M1.7 |
| Average shock speed (km/s) | 455 | 608 | 556 |
| Average shock transit speed (km/s) | 629 | 851 | 775 |
| Average ejecta speed (km/s) | 446 | 572 | 527 |
| Average Alfvenic Mach number | 2.6 | 3.4 | 3.2 |

[a]One shock had a radio data gap, so it was not possible to say whether it was radio loud or not.